\documentclass[twocolumn]{aastex63}

\usepackage{color}
\usepackage[multidot]{grffile} 
\usepackage{amsmath}
\usepackage{array}
\usepackage{xfrac}
\usepackage{multirow}

\usepackage{chngcntr}

\usepackage[bottom]{footmisc}
\usepackage{lipsum}
\usepackage{tablefootnote}

\bibpunct[, ]{(}{)}{;}{a}{,}{,}
\begin{document} 

\title{Open Cluster Chemical Homogeneity Throughout the Milky Way} 

\author{Vijith~Jacob~Poovelil}
\affiliation{Department of Physics \& Astronomy, University of Utah, Salt Lake City, UT, 84112, USA}

\author{G.~Zasowski}
\affiliation{Department of Physics \& Astronomy, University of Utah, Salt Lake City, UT, 84112, USA}

\author{S.~Hasselquist}
\affiliation{NSF Astronomy and Astrophysics Postdoctoral Fellow}
\affiliation{Department of Physics \& Astronomy, University of Utah, Salt Lake City, UT, 84112, USA}

\author{A.~Seth}
\affiliation{Department of Physics \& Astronomy, University of Utah, Salt Lake City, UT, 84112, USA}

\author{John~Donor}
\affiliation{Department of Physics \& Astronomy, Texas Christian University, Fort Worth, TX, 76129, USA}

\author{Rachael~L.~Beaton}
\affiliation{Department of Astrophysical Sciences, Princeton University, 4 Ivy Lane, Princeton, NJ 08544, USA}

\author{K.~Cunha}
\affiliation{Steward Observatory, University of Arizona, Tucson, AZ, 85721, USA}
\affiliation{Observat\'{o}rio Nacional, 20921-400 So Crist\'{o}vao, Rio de Janeiro, RJ, Brazil}

\author{Peter~M.~Frinchaboy}
\affiliation{Department of Physics \& Astronomy, Texas Christian University, Fort Worth, TX, 76129, USA}

\author{D. A. Garc\'{i}a-Hern\'{a}ndez}
\affiliation{Instituto de Astrof\'{i}sica de Canarias (IAC), E-38205 La Laguna, Tenerife, Spain}
\affiliation{Universidad de La Laguna (ULL), Departamento de Astrof\'{i}sica, E-38206 La Laguna, Tenerife, Spain}

\author{K.~Hawkins}
\affiliation{Department of Astronomy, The University of Texas at Austin, Austin, TX, 78712, USA}

\author{K.~M.~Kratter}
\affiliation{Steward Observatory, University of Arizona, Tucson, AZ, 85721, USA}
\affiliation{Department of Astronomy, University of Arizona, Tucson, AZ, 85721, USA}

\author{Richard~R.~Lane}
\affiliation{Instituto de Astronom\'ia y Ciencias Planetarias, Universidad de Atacama, Copayapu 485, Copiap\'o, Chile}

\author{C.~Nitschelm}
\affiliation{Centro de Astronom{\'i}a (CITEVA), Universidad de Antofagasta, Avenida Angamos 601, Antofagasta 1270300, Chile}

\shortauthors{Poovelil et al.}

\begin{abstract}

The chemical homogeneity of surviving stellar clusters contains important clues about interstellar medium (ISM) mixing efficiency, star formation, and the enrichment history of the Galaxy. Existing measurements in a handful of open clusters suggest homogeneity in several elements at the 0.03~dex level. Here we present (i) a new cluster member catalog based only on APOGEE radial velocities and {\it Gaia}-DR2 proper motions, (ii) improved abundance uncertainties for APOGEE cluster members, and (iii) the dependence of cluster homogeneity on Galactic and cluster properties, using abundances of eight elements from the APOGEE survey for ten high-quality clusters. We find that cluster homogeneity is uncorrelated with Galactocentric distance, $|Z|$, age, and metallicity. However, velocity dispersion, which is a proxy for cluster mass, is positively correlated with intrinsic scatter at relatively high levels of significance for [Ca/Fe] and [Mg/Fe]. We also see a possible positive correlation at a low level of significance for [Ni/Fe], [Si/Fe], [Al/Fe], and [Fe/H], while [Cr/Fe] and [Mn/Fe] are uncorrelated. The elements that show a correlation with velocity dispersion are those that are predominantly produced by core-collapse supernovae (CCSNe). However, the small sample size and relatively low correlation significance highlight the need for follow-up studies. If borne out by future studies, these findings would suggest a quantitative difference between the correlation lengths of elements produced predominantly by Type~Ia SNe versus CCSNe, which would have implications for Galactic chemical evolution models and the feasibility of chemical tagging.

\end{abstract}

\section{INTRODUCTION}
\label{sec:intro}
The chemical composition of stars that we see today is a consequence of a sequence of past enrichment events that polluted the interstellar medium (ISM). 
Hence, by studying stellar chemistry, we can learn how these events contributed to enrich the ISM and improve our understanding about the evolution of the Galaxy over time. Open clusters (OC) are particularly interesting objects since they consist of stars that were born together from the same initial molecular cloud, and hence are believed to be chemically homogeneous. Studying the chemistry of these objects can help us trace the ISM pollution rates of different nucleosynthetic processes and ISM mixing efficiency in different locations in the Galaxy.
One must also rely on the assumption of chemical homogeneity of OCs to identify common birth sites using only the chemical signatures of stars, also called chemical tagging \citep{Freeman_2002_chemtag}. Measuring the level of homogeneity of OCs and understanding the factors that can affect it is crucial for the feasibility of chemical tagging.

Cluster chemical homogeneity has been studied in many globular clusters (GCs) and a few OCs. GCs are observed to have inhomogeneities and anti-correlations in most of the light elements~\citep[e.g.,][]{Carretta_2010_GC_anti_correlation,Milone_2018_GC_inhomogeneity,Meszaros_2020_GC_anti-correlations}. Chromosome maps show light element abundance scatter in GCs with masses down to $10^4$~M$_{\odot}$ \citep{Saracino_2019_GC_inhomogeneity_mass}. Heavy element abundance variations are also observed, but only in a small number of massive GCs~\citep[e.g.,][]{Gratton_2020_GC_review}.

The chemistry of OCs, on the other hand, has not been as widely explored, and questions remain regarding the level --- or even presence of --- intrinsic chemical scatter. The Hyades is a well-studied cluster that has been argued to be chemically homogeneous \citep{De_Silva_2006_Hyades,De_Silva_2011_Hyades}, although other recent work on the same cluster identified abundance variations of around 0.02~dex \citep{Liu_2016_Hyades}. M67, another well-studied OC, was found to potentially be inhomogeneous in certain elements from the analysis of two solar twins in the cluster by \citet{Liu_2016_M67}.
However, \citet{Bovy_2016_OC_homogeneity} showed that the scatter of several elemental abundances relative to hydrogen within M67, NGC~6819, and NGC~2420 is as low as 0.03~dex, using APOGEE spectra.
\citet{De_Silva_2007_Collinder_261_homogeneity} demonstrated chemical homogeneity to the 0.05~dex level in seven elements using twelve red giants in the OC Collinder~261,
and \citet[][]{BertranDeLis_2016_O_Fe_scatter} calculated a scatter of [O/Fe] of $\lesssim$0.01~dex in several clusters.

In addition to chemical homogeneity within a single cluster, which is the focus of this paper, the efficacy of chemical tagging also depends on the degree of chemical overlap between different clusters. Intriguingly, some studies have identified chemically indistinguishable pairs of clusters \citep[e.g., NGC~2458 and NGC~2420;][]{Garcia-Dias_2019_NGC_2458-NGC_2420_indistinguishable}, a high degree of overlap in chemical signatures between clusters \citep[e.g.,][]{Blanco-Cuaresma_2015_atomic_diffusion}, and pairs of chemically-indistinguishable stars inside two distinct clusters \citep{Ness_2018_doppelgangers}.

Despite the tension in the literature findings,
OCs are frequently assumed to be chemically homogeneous in cluster studies or for chemical tagging applications. However, there are theoretically and observationally motivated mechanisms that could cause OCs to be inhomogeneous to some level in certain elements.
Dredge-up and atomic diffusion can drive abundance inhomogeneities between cluster members in different stellar evolutionary stages. 
For instance, \citet{Souto_2019_M67_atomic_diffusion} found evidence of atomic diffusion in the open cluster M67, resulting in abundance differences of up to 0.1 dex between stars in different evolutionary states.
\citet{Blanco-Cuaresma_2015_atomic_diffusion} also found variations in chemical signatures for stars belonging to different evolutionary stages within the same cluster. Models of planetary engulfment suggest this process could be responsible for some of the observed elemental scatter in OCs, like the Pleiades \citep{Spina_2018_planetary_engulfment}.

Aside from these mechanisms that can change the observable chemical composition of stars during their lifetime, inhomogeneities could also arise from intrinsic scatter present in the giant molecular cloud before star formation began or significant pollution over the many Myr timescale of the cluster's formation~\citep{Krumholz_2019_Star_clusters_across_cosmic_time}. The correlation length of elements in the initial cloud depends on the series of various enrichment events that produced them and on the efficiency of mixing of the ISM in that region. 
Using hydro-dynamical simulations, \citet{Armillotta_2018_simulationschemtag} found that chemical abundances in OCs should be correlated up to 1 pc, irrespective of the initial correlation lengths of the elements. Thus, given that a typical OC has an effective radius of around 4-5 pc \citep{Kharchenko_2013_catalog}, we may see inhomogeneities in cluster abundances depending on the initial distribution of the elements before star formation began. 

\citet{Bland-Hawthorn_2010_mass_OCs} suggest that all clusters with masses up to $10^4M_{\odot}$, and a significant fraction of those up to $10^5M_{\odot}$, are expected to be homogeneous. However, large star-forming clouds that form more massive clusters, may be subject to pollution from massive stars that become supernovae before star formation is complete. For example, the Sun is suggested to have formed in a cluster with a high-mass star that became a supernova while the Sun was still a protostar~\citep{Looney_2006_Sun_SNe_pollution}.

The many possible mechanisms described above that induce abundance scatter within OCs imply that the level of scatter may depend on properties like the nucleosynthetic groups of elements, range of evolutionary state, cluster size, cluster mass, etc. 
Any systematic difference between the levels of homogeneity for different metals (e.g., between alpha elements and iron-peak elements) may indicate how different enrichment events can affect ISM mixing efficiency. 

Although a few well-studied OCs have been shown to be chemically homogeneous within the observational uncertainties of large scale surveys, there has not been a survey that has looked at a large number of clusters and systematically studied how their chemical homogeneity depends on various Galactic and cluster properties.
So then it is necessary to look at these properties to seek out subtle patterns or behavior that can distinguish among mechanisms.
This paper describes the first such systematic study of OC chemical homogeneity as a function of Galactic and cluster parameters such as Galactocentric distance ($R_{\rm GC}$), vertical height ($|Z|$), age, and mass.

The paper is organised as follows: \S\ref{sec:data} describes the data that we have used from APOGEE (\S\ref{sec:apogee}), Gaia (\S\ref{sec:gaia}), distance catalogs (\S\ref{sec:data_distances}), and cluster catalogs (\S\ref{sec:kharchenko}). We also describe the procedure we use to derive improved uncertainties for APOGEE abundances (\S\ref{sec:uncertainties}). 
\S\ref{sec:membership} explains our kinematics-based cluster membership selection (\S\ref{sec:membership_method}), the validation of our cluster members (\S\ref{sec:validation}), and the final catalog (\S\ref{sec:catalog}). In \S\ref{sec:homogeneity}, we describe how we quantify cluster chemical homogeneity, and
\S\ref{sec:results} contains the analysis of cluster chemical homogeneity versus cluster and Galactic properties. \S\ref{sec:conclusion} summarizes the main results from the paper.

\section{Data}
\label{sec:data}

\subsection{APOGEE}
\label{sec:apogee}

We adopted stellar parameters, chemical abundances, and radial velocities (RVs) from the Apache Point Observatory Galactic Evolution Experiment \citep[APOGEE;][]{Majewski_2017_apogeeoverview}.  APOGEE, one of the component surveys of the Sloan Digital Sky Survey IV \citep[SDSS-IV;][]{Blanton_2017_sdss4}, is a high resolution, near-infrared spectroscopic survey of $\sim$500,000 stars across the Milky Way \citep{Zasowski_2013_targetselection,Zasowski_2017_apogee2targeting}.  Observations are taken with two custom-built, 300-fiber spectrographs \citep{Wilson_2019_apogeespectrographs}, one at the 2.5-m Sloan Telescope at the Apache Point Observatory \citep{Gunn_2006_sloantelescope} and one at the 2.5-m du~Pont Telescope at Las Campanas Observatory \citep{Bowen_1973_duPontTelescope}. We use data from the sixteenth SDSS data release \citep[DR16;][]{Ahumada_2019_SDSSDR16}. 

The pipelines that reduce the data and derive RVs are described in \citet{Nidever_2015_apogeereduction}, and the APOGEE Stellar Parameters and Chemical Abundances Pipeline (ASPCAP) is detailed in \citet{GarciaPerez_2016_aspcap}. The DR16 ASPCAP values were derived by optimizing the comparison of the APOGEE spectra with synthetic spectra computed with Turbospectrum~\citep{Plez_2012_turbospectrum} and MARCS model atmospheres~\citep{Gustafsson_2008_MARCS_model_atmospheres}. The optimization is carried out using the FERRE code~\citep{AllendePrieto_2006_FERRE_code}. The description of the APOGEE data products and abundance reliability for DR14 are presented in \citet[][]{Holtzman_2018_dr13dr14apogee} and \citet{Jonsson_2018_dr13dr14abundances}, while those for DR16 are described in \citet{Jonsson_2020_DR16_data}.

Because our cluster membership determination (\S\ref{sec:membership}) is based solely on kinematic properties, initially we only require reliable kinematical measurements: RVs (discussed here) and PMs (\S\ref{sec:gaia}).  Additional cuts on data quality used for the analysis in \S\ref{sec:results} are described \S\ref{sec:results_elements_members}.
To ensure reliable kinematic cluster member selection, we restrict our sample to have APOGEE RV uncertainties (VERR) of $<$0.1~km~s$^{-1}$.
We also remove a small number of stars with implausibly large velocities by requiring $\rm |VHELIO\_AVG| < 5000$~km~s$^{-1}$.

To remove potential binaries from our sample, which may inflate the characteristic velocity signatures inferred for our clusters, we remove stars with visit-to-visit RV variations (VSCATTER) $>$1~km~s$^{-1}$ \citep{Price-Whelan_2020_binaries,Badenes_2018_binaries} and stars found in the binary catalog of \citet{Price-Whelan_2020_binaries}.
Given that the time baseline for the majority of APOGEE sources is less than a year, APOGEE is sensitive to detect binaries with periods less than a few years and separated by distances less than a few AU \citep{Price-Whelan_2020_binaries}. We note that since the VSCATTER limit that we impose is more sensitive to massive binaries, which tend to be more centrally concentrated, and that binary properties are correlated with metallicity \citep[e.g.,][]{Moe_2019_binarymetallicity,Badenes_2018_binaries}, it is conceivable that these binary rejection cuts could induce some metallicity-dependent spatial sampling patterns. We confirmed that this is not the case in our sample because this limit removes a tiny fraction of stars ($<$5\%), whose spatial distributions are indistinguishable from those that pass these cuts. 

\subsection{Re-derived Abundance Uncertainties}
\label{sec:uncertainties}

\subsubsection{Motivation and outline}
Earlier studies have highlighted the possibility that the uncertainties for some of the APOGEE abundances in earlier data releases are overestimated \citep[e.g.,][]{Ness_2018_doppelgangers}, and our own preliminary analyses of the abundance dispersions in our clusters supported this assessment. Prior to DR16, uncertainties on the abundances in the APOGEE data releases were determined by examining the spread of abundances in well-sampled clusters \citep[][]{Holtzman_2018_dr13dr14apogee}, which required assuming that the cluster had no intrinsic spread. This assumption makes the uncertainties unsuitable for the study of chemical homogeneity.

To address this, we recalculate the random abundance uncertainties for the stars in our analysis using an improved method\footnote{We emphasize that this empirical procedure for determining the uncertainties in the abundances accounts only for the random component of the uncertainties, which is essential for deriving the intrinsic abundance scatter in a cluster. Any systematic components, such as stellar parameter-dependent abundance variations due to departures from LTE or hydrostatic equilibrium, or systematics in the atomic data or instrumental distortions, are not captured by this procedure.}. We worked closely with the APOGEE team to subsequently adapt our approach into the DR16 uncertainty determination. The approach used in DR16~\citep{Jonsson_2020_DR16_data} relies on a parametric fit to calculate the uncertainties for all the APOGEE stars. In this paper, however, we adopt a non-parametric approach because we find that a simple analytic function cannot adequately capture the relationships between uncertainty and SNR, T$_{\rm eff}$, and [M/H]. 

In brief, we use the differences in [X/Fe] values derived by ASPCAP for independent spectra of the same star to compute a relationship between the abundance uncertainties and SNR, $T_{\rm eff}$, and [M/H] (\S\ref{sec:uncertainties_calculate}). Then, we use this relationship to compute the uncertainties for our cluster members (\S\ref{sec:uncertainties_assign}).

\subsubsection{Calculating uncertainties using multiple visits}
\label{sec:uncertainties_calculate}

In general, the ASPCAP pipeline is run on the stacked spectrum of each star, which comprises all visits to that star. For a subset of APOGEE stars, however, ASPCAP is run on the individual visit-level spectra, providing multiple independent sets of stellar parameter and abundance measurements for single stars\footnote{These measurements are contained in the ``allCal'' file as part of APOGEE's data releases.}. We use these sets to estimate the random uncertainty of the ASPCAP measurements as a function of SNR, T$_{\rm eff}$, and [M/H]. The variations between ASPCAP values for spectra of the same star, at the same SNR, provides a more realistic representation of the random measurement uncertainties than the ones derived from cluster dispersions in earlier APOGEE DRs. 

We define a sample of stars, hereafter called the Uncertainty Training (UT) sample, with ASPCAP solutions derived from two or more visit spectra with similar SNR ($\Delta$SNR/SNR $\le$ 20\%), resulting in similar temperature ($\Delta\!T_{\rm eff} \le 100$ K) and metallicity ($\rm \Delta [M/H] \le 0.07$~dex) values. These similarity criteria are imposed to ensure that differences in [X/Fe] are not due to different global spectral fits. We restrict our analysis to giant stars (using the ASPCAP\_CLASS column and a limit of $\log{g}<3$).

The final UT sample of 8729 stars is then divided into five bins of SNR: 50--70, 70--100, 100--130, 130--200, and $>$200.  These bins are chosen to provide finer sampling at lower SNR, where the effect of SNR on output parameters is larger; above $\rm SNR \approx 150$, there is a weak relationship with uncertainties. We explored dividing the sample into finer bins in SNR. However, there were no significant changes in the final derived uncertainties, and there was an increased risk of undersampling each bin so that the estimated uncertainty would not capture the effect of the varying range of stellar parameters. 

We adopt a Voronoi binning scheme in $T_{\rm eff}$--[M/H], within these fixed SNR ranges, to ensure both reliable measurements of the uncertainties in less populated regions of the parameter space, and high resolution measurements where possible. The python package {\it vorbin} \citep{Cappellari_2003_vorbin} is used to group the UT sample into 2D bins of $T_{\rm eff}$ and [M/H], targeting at least 30 stars per bin. The final bins are populated with between nine and 69 stars per bin, with an average of around 33. 

The differences between pairs of visit-level [X/Fe] values for individual stars can be used to compute the standard deviation of the distribution from which the pairs were originally drawn. We assume this distribution to represent the intrinsic ASPCAP random uncertainty.
The quantities are related by

\begin{equation}
\label{eqn:metric}
   \rm e_{[X/Fe],k} = \frac{\sqrt{\pi}}{2} median(|[X/Fe]_i - [X/Fe]_j|),
\end{equation}
where $e_{\rm [X/Fe],k}$ is the abundance uncertainty associated with the $k^{th}$ bin of $T_{\rm eff}$ and [M/H] at a given SNR, and ${\rm [X/Fe]}_i$ and ${\rm [X/Fe]}_j$ refer to abundance measurements derived from two independent visit spectra of the same star. The top row of Figure~\ref{fig:uncertainty_examples} shows the distribution of UT stars in [M/H] and $T_{\rm eff}$, in bins of SNR, colored by the $e_{\rm [Mg/Fe],k}$ values. This demonstrates the complex pattern of $e_{\rm [Mg/Fe],k}$ derived in this way and the difficulty of describing such behavior with simple analytical expressions. Similar figures for all elements we analyze in~\S\ref{sec:results} can be found in Appendix~\ref{sec:app_uncertainties}.

We looked for potential dependencies of $e_{\rm [X/Fe],k}$ on [X/Fe] itself --- e.g., that stars with enhanced [Mg/Fe] may have different uncertainties than stars with solar [Mg/Fe] at the same $T_{\rm eff}$, [M/H], and SNR --- by deriving $e_{\rm [X/Fe],k}$ in the same manner as above, but separately for stars with high and low [Mg/Fe].  We found no significant differences in the $e_{\rm [X/Fe],k}$($T_{\rm eff}$, [M/H], SNR) patterns. We also found no difference in the results discussed in \S\ref{sec:results} when using uncertainties derived from the solar [Mg/Fe] part of the UT sample as compared to using the full sample. 

\subsubsection{Assigning uncertainties to stars}
\label{sec:uncertainties_assign}

Given the computed array of $e_{\rm [X/Fe],k}$ for each bin of $T_{\rm eff}$, [M/H], and SNR, we can then sort any other star into a bin and assign it $e_{\rm [X/Fe],k}$ values. We perform this sorting by training a Gaussian Naive Bayes classifier algorithm \citep{scikit-learn} on the UT sample, and then applying the trained classifier to our cluster member sample (\S\ref{sec:catalog}). This results in a bin membership probability for each star; we assign each star an uncertainty summed from all of the Voronoi bins and weighted by each bin's probability for that star:
\begin{equation}
\label{eqn:probabilities}
    e_{[X/Fe]} = \sum_k p_k e_{[X/Fe], k},
\end{equation}
where $k$ is the bin index, $p_k$ is the probability of bin $k$ for this star, and $e_{\rm [X/Fe]}$ is the final uncertainty for the star. Adopting this weighted average uncertainty ensures that all stars falling within a range of stellar parameters will not be assigned identical values for their abundance uncertainties due solely to the binning scheme. This approach also smooths the transitions between the bin edges. Ultimately, the uncertainties assigned to a star with stellar parameters near a bin edge are not driven by the distribution of the bins, but by the uncertainties estimated from the UT sample that fall within a neighboring area of the star in [M/H] and $T_{\rm eff}$.

\begin{figure*}[!hbt]
\includegraphics[angle=0, clip, width=1.0\textwidth]{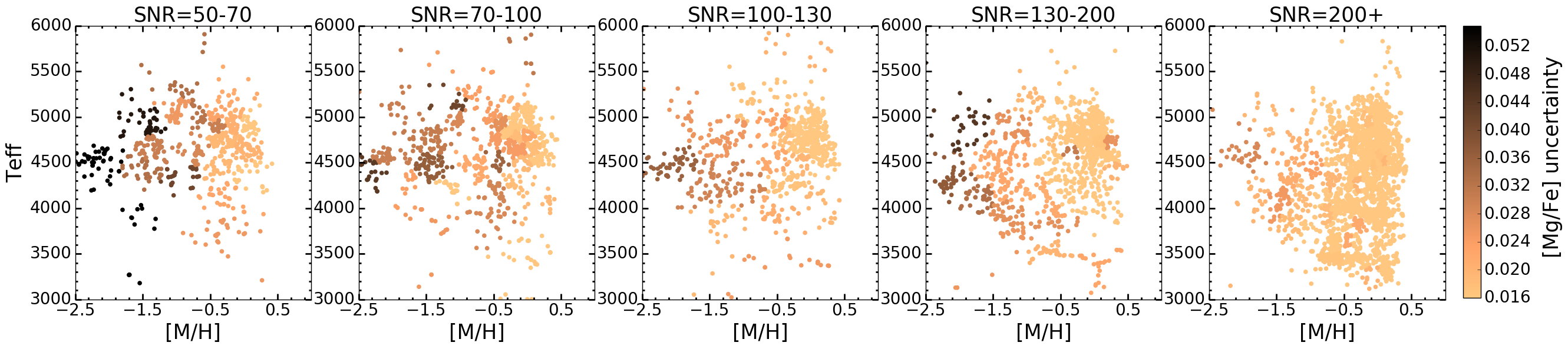}
\includegraphics[angle=0, clip, width=1.0\textwidth]{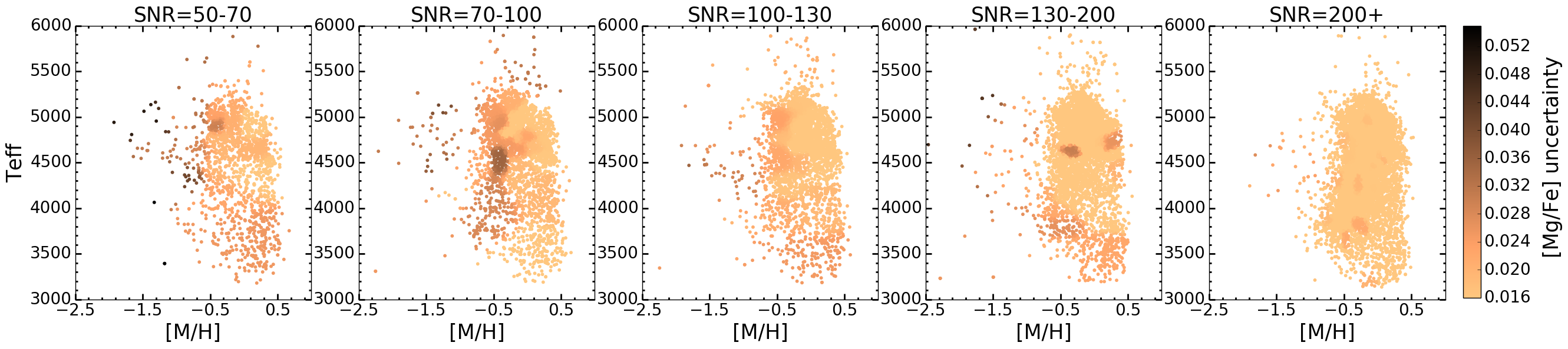}
\includegraphics[angle=0, clip, width=1.0\textwidth]{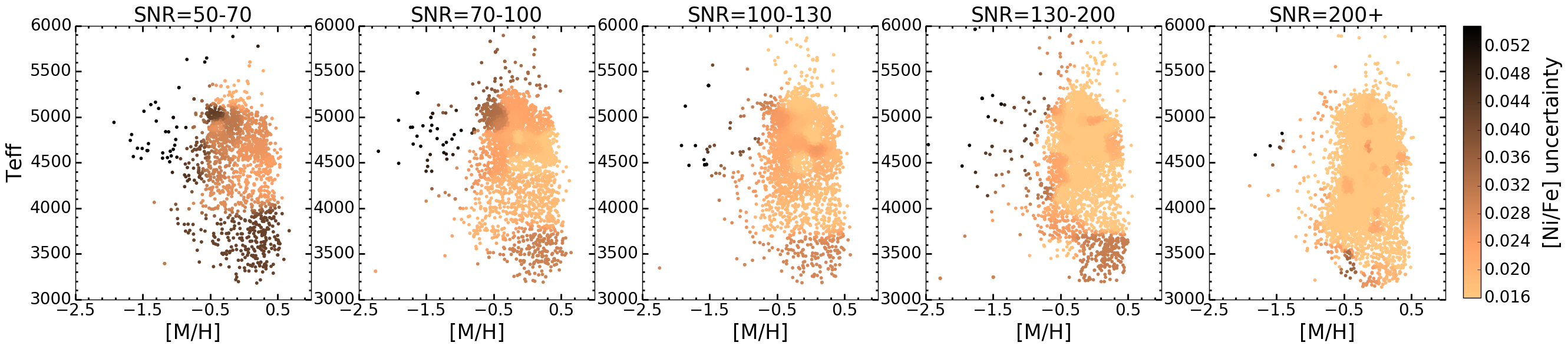} \includegraphics[angle=0, clip, width=1.0\textwidth]{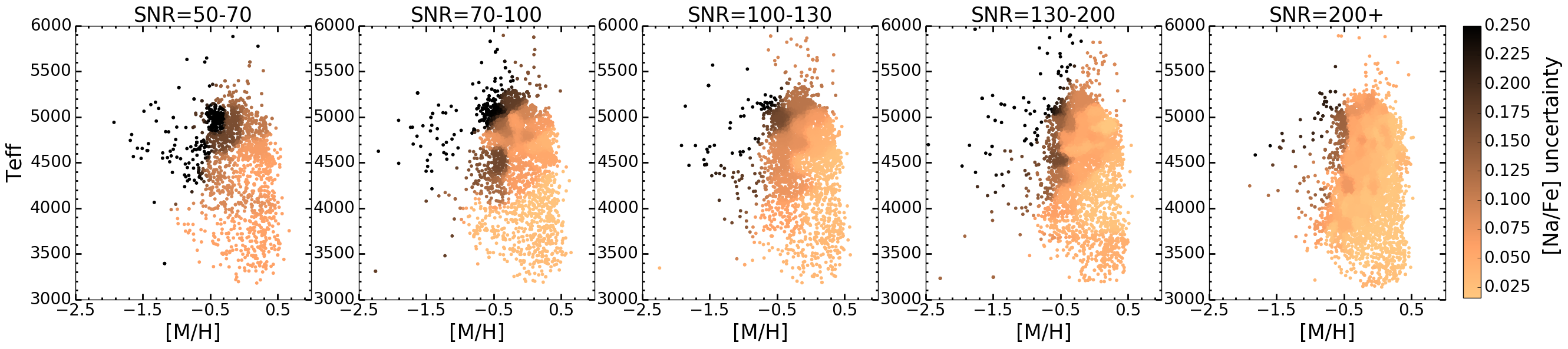} 
\caption{
  First row: the Uncertainty Training (UT) sample, divided by SNR and Voronoi-binned by $T_{\rm eff}$--[M/H], as described in the text (\S\ref{sec:uncertainties}). The stars in each bin are colored by the computed $e_{\rm [X/Fe]}$ for that bin; [Mg/Fe] is shown here as an example. Note that many adjacent bins have nearly identical $e$ values and are indistinguishable in this representation.
  Second row: Weighted uncertainties for [Mg/Fe] of the cluster member sample (\S\ref{sec:membership}). The uncertainties assigned to the cluster members trace the pattern seen in first row.
  Third row: Weighted uncertainties for [Ni/Fe] of the cluster member sample.
  Fourth row: Weighted uncertainties for [Na/Fe] of the cluster member sample.  Note that the range in the colorbar has been increased.
  }
\label{fig:uncertainty_examples}
\end{figure*}

The bottom 3 rows of Figure~\ref{fig:uncertainty_examples} show examples of these rederived uncertainties for [Mg/Fe], [Ni/Fe], and [Na/Fe] (note the differences in color scaling) along [M/H] and $T_{\rm eff}$, in bins of SNR. All show the expected improvement in precision with higher SNR. Other generic patterns are also clearly visible --- for example, the increase in uncertainty at low metallicities and/or high temperatures, where lines become weaker, and at very low temperatures, where lines become increasingly blended. Mg and Ni have lower uncertainties compared to Na, which is expected due to the difficulty in measuring Na lines in APOGEE. Each element also has its own unique patterns, reflecting the range of difficulty in measuring lines of different elements in different parts of stellar parameter space.

The DR16 uncertainties are not systematically higher or lower than the uncertainties derived here for stars in our cluster sample (\S\ref{sec:membership}), and we find qualitatively similar results for the analysis in \S\ref{sec:results} if we use either set of values.

\subsection{\it Proper Motions}
\label{sec:gaia}

For the cluster membership selection in \S\ref{sec:membership}, we use proper motions (PMs) from DR2 of the {\it Gaia} mission \citep{Gaia_collaboration_2018}. We require that the errors in the proper motion measurements be smaller than 2.0~mas~yr$^{-1}$ and the renormalized unit weight error (RUWE) be less than 1.4 \citep{Ziegler_2020_Gaia_RUWE}. 
In addition, cluster-specific limits are imposed on the spatial distribution, magnitude, and color of the stars, as described in \S\ref{sec:PMs}.

\subsection{Stellar Distances}
\label{sec:data_distances}

For our analysis in \S\ref{sec:results}, we use spectrophotometric distances calculated using the method described in \citet[][RA17]{Rojas-Arriagada_2017_APOGEE_distances}. We also use the StarHorse \citep{Queiroz_2018_StarHorse,Queiroz_2020_StarHorse} and astroNN \citep{Leung_2019_astroNN} distances to compare with our RA17 estimates. The results of this paper are not affected by the choice of the distance catalog used to determine cluster distances. Further discussion can be found in \S\ref{sec:results}.

\subsection{Literature Cluster Parameters}
\label{sec:kharchenko}

We adopt the Milky Way Star Clusters catalog \citep[][hereafter K13]{Kharchenko_2013_catalog} as the base catalog for our membership search \S\ref{sec:membership}. 
We use K13 cluster center coordinates and angular radii to define the search limits, and we consider the cataloged distances, ages, and metallicities (in conjuntion with APOGEE-derived values) when assessing our membership selection. 
From the total sample of 3208 clusters in K13, we only consider the 366 clusters that have six or more APOGEE stars (\S\ref{sec:apogee}) within their cluster radii (\S\ref{sec:membership_cluster_coordinates}). We narrow our cluster sample to the ten most-populated high-quality clusters for our analysis in \S\ref{sec:results}.

\section{Cluster Membership}
\label{sec:membership}

The first step of our open cluster analysis is to identify cluster members in the APOGEE sample. Numerous methods have been demonstrated in the literature, typically adopting some combination of RVs, proper motions, metallicities, and position in the CMD
\citep[e.g.,][]{Frinchaboy_2008_OC_membership_method,Meszaros_2013_OC_membership,Donor_2018_occam2}.
As we are interested in the chemical homogeneity of the clusters, we design our membership selection around kinematical information only: RVs from APOGEE (\S\ref{sec:apogee}) and proper motions from {\it Gaia}-DR2 (\S\ref{sec:gaia}). This is similar in spirit to the approach taken by \citet{CantatGaudin_2018_GaiaOCs}, hereafter C18, who use {\it Gaia}-DR2 information only. 

\begin{figure}[hbt!]
\includegraphics[angle=0, clip, width=0.49\textwidth]{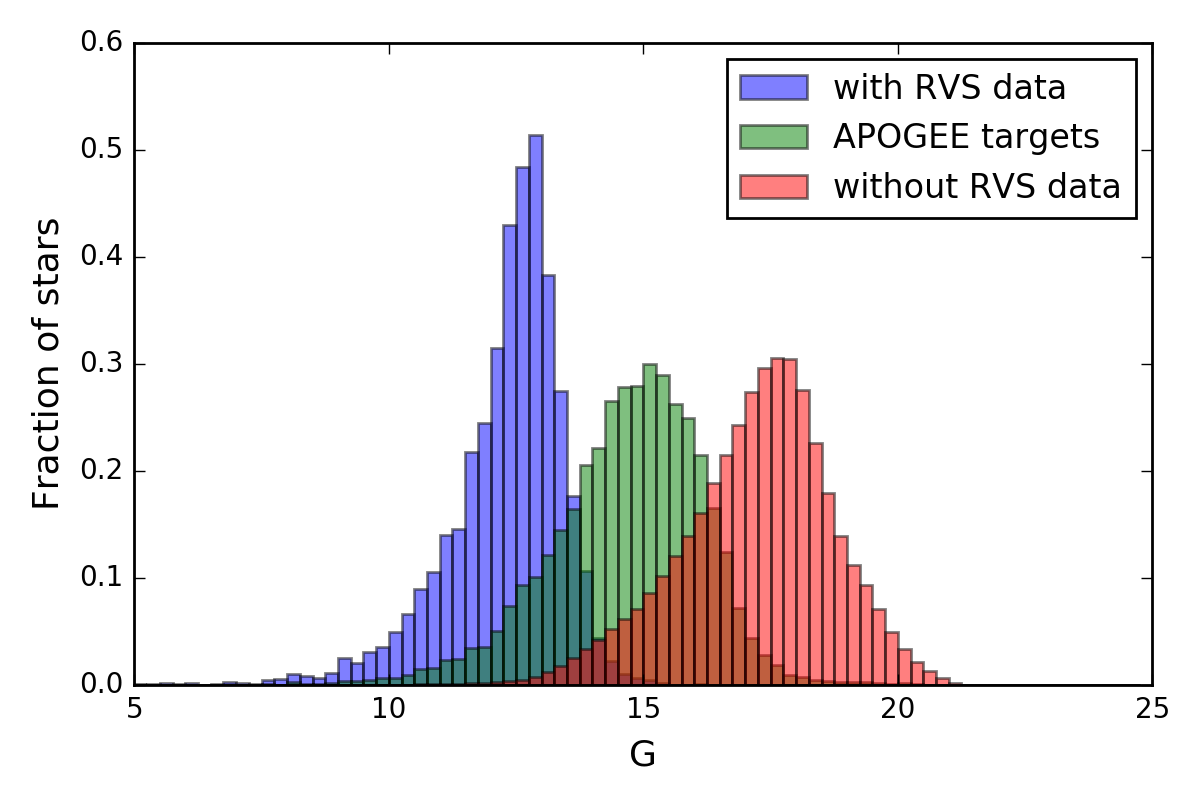}
\caption{
{\it Gaia}-DR2 $G$-band magnitude of stars with (blue) and without (red) {\it Gaia}/RVS data (normalized separately), compared to the APOGEE stars (green) belonging to our clusters. Since \textit{Gaia}-DR2 only has RVs for stars with $G \lesssim 13$, and most of our clusters have RGB stars fainter than this, we do not use \textit{Gaia}/RVS for the RVs in our selection.
} 
\label{fig:Gaia_g-band}
\end{figure}

Figure \ref{fig:Gaia_g-band} shows the distribution of the {\it Gaia}-DR2 $G$-band magnitude of stars with (blue) and without (red) Gaia/RVS data, compared to the APOGEE stars (green) in the vicinity of several of our cluster candidates.
This figure highlights why APOGEE RVs are necessary for the objects in our sample; due to a combination of distance and extinction, most of our stars are too faint to have {\it Gaia}-DR2 RVS radial velocities.

\subsection{Method}
\label{sec:membership_method}

Figures~\ref{fig:membership_NGC6819}--\ref{fig:membership_ASCC116} demonstrate the procedure described below for a well-studied open cluster (NGC~6819), a poorly-studied cluster (FSR~0494), and a K13 cluster not recovered by our membership method (ASCC~116), respectively.

\subsubsection{Cluster Coordinates}
\label{sec:membership_cluster_coordinates}

For each cluster, we start our membership search with APOGEE and {\it Gaia}-DR2 stars within twice the cluster radius, 2$R_{\rm cluster}$, using central coordinates and ``total'' cluster radii (their $r_2 = R_{\rm cluster}$) from K13 (\S\ref{sec:kharchenko}). The ``central stars'' (within 1$R_{\rm cluster}$) define the cluster's kinematical signature, and the ``annulus stars'' (between 1.5--2$R_{\rm cluster}$) define the background distribution (e.g., Figure~\ref{fig:membership_NGC6819}a-b). We only consider the 366 K13 clusters that have six or more stars within 1$R_{\rm cluster}$ in the APOGEE catalog that meet the quality criteria above.

\subsubsection{Radial Velocities}
\label{sec:RVs}

We search for RV peaks associated with co-moving stars in each cluster (e.g., Figure~\ref{fig:membership_NGC6819}c) location by subtracting a kernel density estimate (KDE) of the annulus stellar RV distribution (shown in green) from that of the central stars (shown in blue). The residual (shown in red) is then fit with a Gaussian to determine the central RV ($\langle {\rm RV} \rangle$) and the width ($\sigma_{\rm RV}$) of the dominant peak. We also measure the ratio of the Gaussian amplitude ($A_{\rm RV}$) to the standard deviation of the residuals ($\sigma_{\rm resid}$) more than 3$\sigma_{\rm RV}$ away from the Gaussian center; this metric quantifies the strength of the RV signal. Visual inspection demonstrates that for our clusters, an $A_{\rm RV}/\sigma_{\rm resid} > 9.5$ corresponds to a cluster in RV space. Smaller values tend to be dominated by noisy residuals driven by a low number of annulus stars.

\subsubsection{Proper Motions}
\label{sec:PMs}

We obtain proper motion information for all stars within 2$R_{\rm cluster}$ using the \textit{Gaia} TAP+ query from the astroquery package in python and keep stars that pass the quality cuts mentioned in \S\ref{sec:gaia}. 

Next, we ensure that the {\it Gaia}-DR2 data sample has the same color-magnitude range as the APOGEE stars by matching to the `apogeeObject' files used in the APOGEE targeting pipeline, which contain the 2MASS \citep{Skrutskie_06_2mass}, {\it Spitzer}--IRAC GLIMPSE \citep{Benjamin_05_glimpse,Churchwell_09_glimpses}, and AllWISE \citep{Wright_10_WISE,Cutri_2013_allwise} photometry used to calculate extinction \citep{Majewski_11_RJCE,Zasowski_2013_targetselection,Zasowski_2017_apogee2targeting}.
We then restrict the {\it Gaia}-DR2 stars to the same $(J-K_s)_0$ and $H$ limits sampled by the APOGEE stars in the vicinity of that cluster (generally $(J-K_s)_0 \ge 0.5$ and $7 \le H < 12.2$). 

We use this cut to ensure that the PM distribution obtained from {\it Gaia}-DR2 stars is an accurate representation of the APOGEE stars that we are considering for membership. However, we compare the membership with and without using this cut for all the ten clusters studied in \S\ref{sec:results}. Although the PM distribution is altered slightly, the final cluster members determined are the same for these clusters irrespective of the color-magnitude cut.

In a small fraction of cases (6\%), the apogeeObject files do not span the full background annulus region, but we have confirmed that the distributions of $\mu_{\rm RA}$ and $\mu_{\rm Dec}$ do not change across the small angular scales of our clusters, so we consider even these partial annuli to be representative of the background distributions.

As with the RVs (\S\ref{sec:RVs}), we compare KDEs of the central and annular distributions to identify any signal of the cluster, this time with 2D KDEs ($\mu_{\rm RA} \times \mu_{\rm Dec}$, shown in Figure~\ref{fig:membership_NGC6819}d). Because the annular PM distribution is much less noisy than in case of RV, we model the entire central PM distribution (shown in blue) as the sum of a 2D Gaussian and a scaled copy of the annular PM distribution (shown in green). The best-fit Gaussian (shown in red) center ($\langle {\rm PM} \rangle$) and 2D dispersion ($\sigma_{\rm PM}$(RA, DEC)) are taken as the PM distribution of the co-moving cluster stars.

\subsubsection{Computing Membership Probabilities}
\label{sec:probabilities}

\begin{figure}[hbt!] 
\includegraphics[angle=0, clip, width=0.49\textwidth]{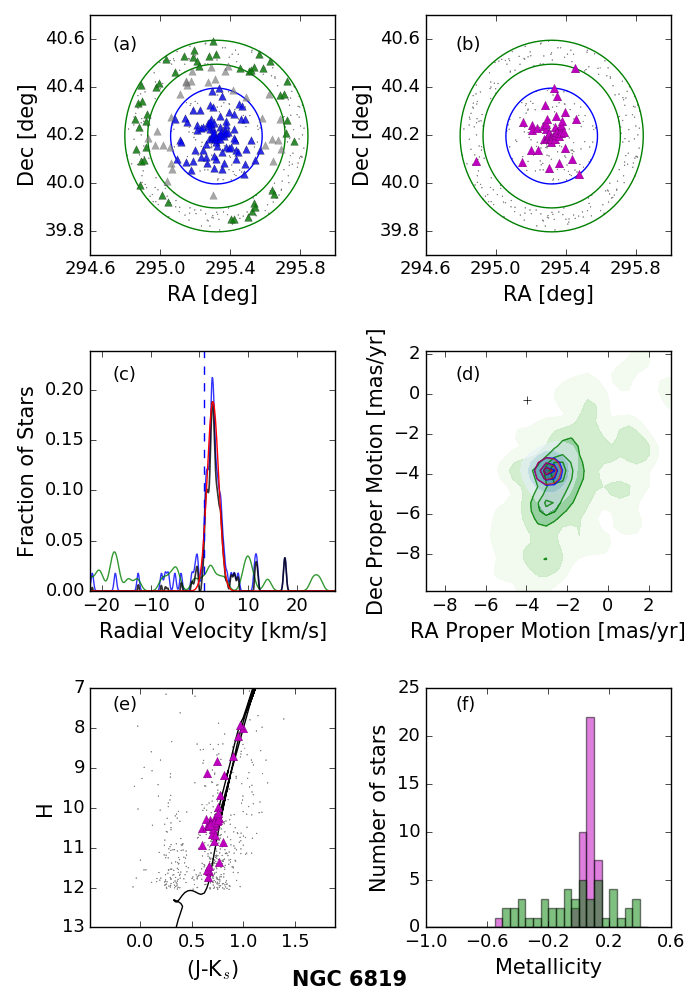}
\caption{
Proof-of-concept membership selection for NGC 6819: APOGEE stars within annulus and central regions are shown in green and blue, respectively, while final cluster members are shown in purple. (a) and (b): Stellar distribution in RA and DEC. APOGEE and \textit{Gaia}-DR2 stars are shown as triangles and points, respectively. (c) and (d): Distribution of RVs and PMs. Fits for the subtracted distributions in RV and PM are shown in red. Diagnostic plots for final cluster members: (e) Color-magnitude diagram along with a Padova isochrone corresponding to the cluster; (f) Metallicity distribution of cluster stars as compared to annulus stars. See text for details.
}
\label{fig:membership_NGC6819}
\end{figure}

\begin{figure}[hbt!]
\includegraphics[angle=0, clip, width=0.49\textwidth]{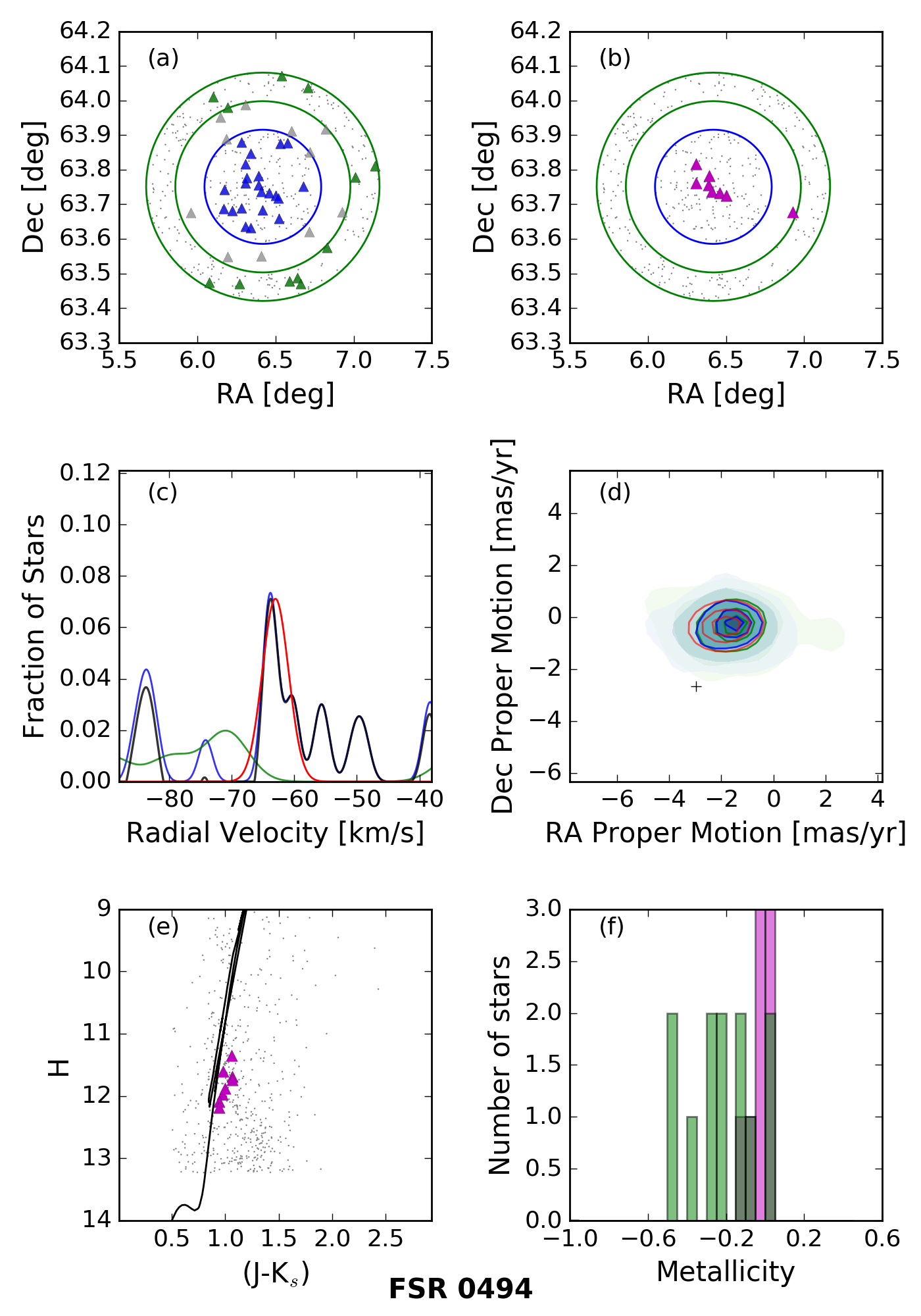} 
\caption{
Same as Figure \ref{fig:membership_NGC6819}, but for FSR 0494, a lesser studied OC.
} 
\label{fig:membership_FSR0494}
\end{figure}

\begin{figure}[hbt!] 
\includegraphics[angle=0, clip, width=0.49\textwidth]{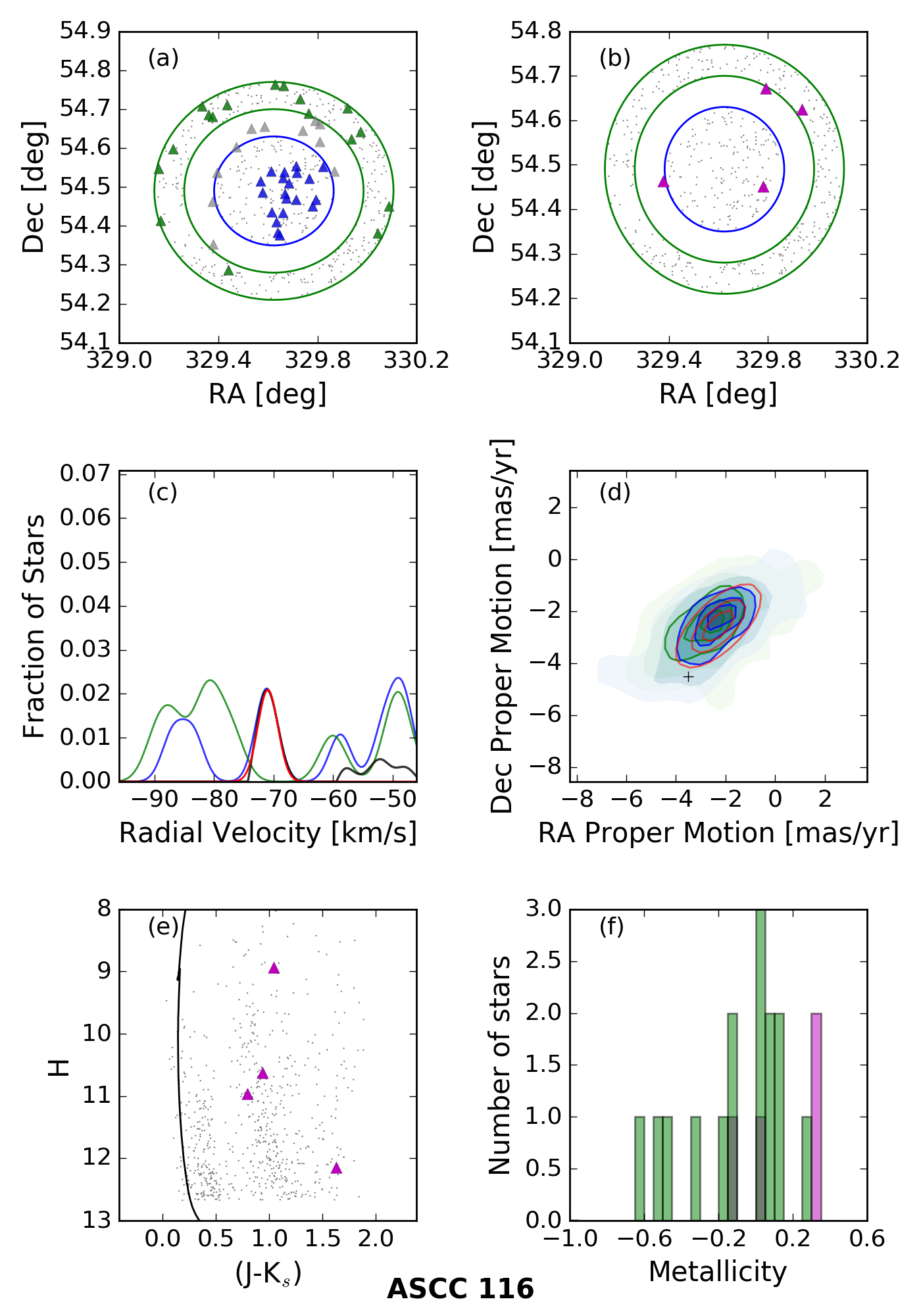}
\caption{
Same as Figure \ref{fig:membership_NGC6819}, but for ASCC 116, shown here as an example of a cluster where the diagnostic plots do not confirm the presence of an OC.
}
\label{fig:membership_ASCC116}
\end{figure}

We first compute cluster membership probabilities, based on RVs and PMs, for each star within 2$R_{\rm cluster}$ of a cluster. These probabilities are the values of Gaussian distributions with the means and standard deviations derived from the RV and PM fitting in \S\ref{sec:RVs}--\S\ref{sec:PMs}, scaled to have a maximum value of unity. 
We consider as likely members stars that fall within a 3$\sigma$ window (shown in purple in Figures~\ref{fig:membership_NGC6819}--\ref{fig:membership_ASCC116}) on the combined RV--PM probability. We give equal weighting to each kinematic dimension while calculating the combined probability, since weighting each dimension on how distinct it is from the background did not yield any changes in the final selected cluster members. 
We choose a selection window of 2$\sigma$ on the combined probability for the analysis in \S\ref{sec:results} since we observed a few outliers in the metallicity distribution (e.g., Figure~\ref{fig:membership_NGC6819}f) that were removed when we used a stricter cut.

Examples of this entire procedure and its results are demonstrated in Figures~\ref{fig:membership_NGC6819}--\ref{fig:membership_ASCC116}. 
The color coding for all panels is as follows: locations and kinematical information for the central APOGEE stars are plotted in blue, for the annulus APOGEE stars in green, for the final members in purple, and for the {\it Gaia}-DR2 stars used for the PM background in gray points. The top row (panels a and b) show the on-sky distribution of stars, with the inner $R_{\rm cluster}$ in a blue circle and the outer annulus enclosed in green circles at 1.5 and 2$R_{\rm cluster}$. Panel~a shows the stars used in characterizing the cluster (\S\ref{sec:RVs}--\S\ref{sec:PMs}), and panel~b shows the final cluster members (\S\ref{sec:probabilities}). 

The middle row (panels c and d) show the RV and PM distributions of the central and annulus stars, with the fitted residual peaks in red. The bottom row (panels e and f) are not used for membership selection and are only examined when setting reliability flags for entire clusters (\S\ref{sec:validation}). 

Figure~\ref{fig:membership_NGC6819} shows the recovery of the well-studied cluster NGC~6819 \citep[e.g.,][]{Hole_2009_ngc6819rvs,Platais_2013_ngc6819pms,Yang_2013_ngc6819photometry,Wu_2014_keplerclusterRGBs,LeeBrown_2015_ngc6819abundances}, Figure~\ref{fig:membership_FSR0494} shows the recovery of the less-studied cluster FSR~0494 \citep{Froebrich_2007_clustersearch,Zasowski_2013_MIRclusters,Donor_2018_occam2}, and Figure~\ref{fig:membership_ASCC116} shows the non-recovery of the cluster catalogued as ASCC~166 \citep[e.g.,][]{Kharchenko_2005_MWOCs,CantatGaudin_2018_GaiaOCs}.

\subsection{Validation}
\label{sec:validation}

In the membership selection examples in Figures~\ref{fig:membership_NGC6819}--\ref{fig:membership_ASCC116}, the left side of the bottom row (panel e) shows the ($J-K_s, H$) color--magnitude diagram of the cluster members and background stars, along with a shifted PARSEC isochrone \citep{Bressan_2012_parsec,Marigo_2017_parseccolibri} corresponding to the cluster's distance, metallicity, and extinction, either known from K13 or approximated from the cluster members themselves. The right side of the bottom row (panel f) shows the APOGEE metallicity distributions of the annulus stars (green) and the kinematically-selected member stars (purple). We use these two pieces of data when setting reliability flags for clusters in the final catalog (\S\ref{sec:catalog}). Although we do not use
metallicity  in  determining  cluster  members, we use it to flag clusters that do not have a clear and distinct MDF, compared to the background stars, described below.

We classify clusters into ``GOOD'', ``WARN'', ``INSUFFICIENT\_DATA'', and ``UNRECOVERED'' categories based on our confidence in the recovery of genuine cluster members:
\begin{itemize} \itemsep -2pt
    \item GOOD:  Clusters that have distinct kinematics ($A_{\rm RV}/\sigma_{\rm resid} > 9.5$; \S\ref{sec:RVs}) and metallicity dispersion less than 0.13 dex. 
    \item WARN: Clusters that have distinct kinematics ($A_{\rm RV}/\sigma_{\rm resid} > 9.5$; \S\ref{sec:RVs}) and metallicity dispersion greater than 0.13 dex.
    \item INSUFFICIENT\_DATA:  Clusters that have distinct kinematics ($A_{\rm RV}/\sigma_{\rm resid} > 9.5$; \S\ref{sec:RVs}) but fewer than 5 cluster members, making it difficult to interpret the diagnostic CMD and MDF distributions (e.g., Figure~\ref{fig:membership_ASCC116}e-f). 
    \item UNRECOVERED:  Clusters that do not have distinct kinematics ($A_{\rm RV}/\sigma_{\rm resid} < 9.5$; \S\ref{sec:RVs}).
\end{itemize}

We observed that the cluster metallicity distributions are generally either very tight or indistinguishable from the background. By visual inspection of all the clusters, this bifurcation is captured using a metallicity dispersion cut of 0.13 dex, with clusters having a higher metallicity dispersion classified using the ``WARN'' flag. 

Clusters that have too few stars to be validated using the diagnostic plots are included in the ``INSUFFICIENT\_DATA'' classification. Figure~\ref{fig:membership_ASCC116} shows an example of such an object, where our membership method fails to confirm a co-moving object at the location of a K13 cluster. Such cases are expected, since APOGEE uses specific sets of color and magnitude limits and does not target the entire sky homogeneously. We include Figure~\ref{fig:membership_ASCC116} to highlight the robustness of this method and validation to false positives. For the analysis in \S\ref{sec:results}, we only use clusters that have a ``GOOD'' validation flag.

\subsection{Catalog}
\label{sec:catalog}

We generate two catalogs based on the kinematic membership selection above. One contains all of the stars within 2$R_{\rm cluster}$ that meet the membership criteria described in \S\ref{sec:probabilities} for the GOOD, INSUFFICIENT\_DATA, and WARN clusters (\S\ref{sec:validation}).
This includes, for each star, the APOGEE ID and stellar coordinates, the name of and distance (in arcmin) from the center of the cluster to which it belongs, and the number of sigmas from the center of the membership probability distribution (\S\ref{sec:probabilities}) in both RV and PM dimensions. A sample of this table is shown in Table~\ref{tab:cluster_members_sample_table}, which is published in its entirety in machine-readable format online. 

The second catalog contains the properties of the clusters themselves, outlined in Table~\ref{tab:catalog_example_clusters}. For each GOOD, INSUFFICIENT\_DATA, and WARN clusters, we give the central coordinates, radius, and age from K13, along with the average distances and metallicities of the member stars, a suite of kinematic fitting parameters and metrics, and other metadata. Table~\ref{tab:catalog_example_clusters} is published in its entirety in machine-readable format. 
A subsample of Table~\ref{tab:catalog_example_clusters} containing important kinematic and chemical information for the ten OCs used in \S\ref{sec:results} is shown in Table~\ref{tab:subsample_clusters}.
Of the 366 K13 clusters with six or more APOGEE stars within 1$R_{\rm cluster}$, 34 are included with the GOOD flag, 38 have WARN, 11 have INSUFFICIENT\_DATA, and 283 are flagged as UNRECOVERED.

\begin{figure*}[!hptb] 
\includegraphics[angle=0, clip, width=\textwidth]{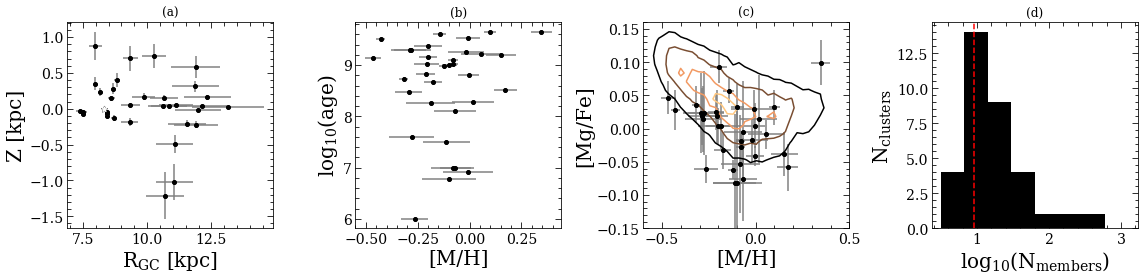} 
\caption{
{\bf Summary of cluster catalog (\S\ref{sec:catalog}).} 
Panel (a) shows the distribution in $R_{\rm GC}$ and $Z$ of clusters flagged as GOOD, using heliocentric distance estimates from the APOGEE member stars. 
Panel (b) shows the distribution of GOOD clusters in [M/H] (from the APOGEE member stars) and log(age) (from K13). 
Panel (c) shows the distribution of GOOD clusters in [M/H] and [Mg/Fe] (against a background of MW stars).
Panel (d) shows the histogram of the number of cluster members in the GOOD clusters. For the analysis in \S\ref{sec:results}, we only use clusters with at least nine members, shown by the vertical dashed line.
}  
\label{fig:catalog_properties}
\end{figure*}

Figure~\ref{fig:catalog_properties} summarizes several properties of the GOOD clusters from our catalog.
Figure~\ref{fig:catalog_properties}a shows the clusters' Galactic $R_{\rm GC}$--$Z$ distribution, and Figure~\ref{fig:catalog_properties}b shows the distribution of their mean [M/H] and $\log({\rm age})$ values. In Figure~\ref{fig:catalog_properties}c, we plot the distribution of mean [M/H] and [Mg/Fe] over a background of APOGEE stars with similar Galactic radius and height ($R_{\rm GC}=5-15$~kpc and $|Z|<2$~kpc), selected using the same quality criteria described in \S\ref{sec:apogee} and \S\ref{sec:results}. Figure~\ref{fig:catalog_properties}d shows a histogram of the number of cluster members identified, with the cutoff of nine members used in \S\ref{sec:results} indicated with a red dashed line.

We find similar member sample sizes as \citet{Donor_2020_occam3}, who also studied APOGEE DR16, for the clusters in common. We calculate a metallicity gradient (using [M/H]) of $-0.096 \pm 0.016$~dex~kpc$^{-1}$ for the sample, spanning $R_{\rm GC}=7-12$~kpc and $|Z_{\rm GC}|<1$~kpc. This value is within the uncertainties of, but slightly steeper than, previous calculations of the metallicity gradient \citep[e.g., $-0.079 \pm 0.005$ and $-0.085 \pm 0.019$~dex~kpc$^{-1}$;][]{Donor_2018_occam2, Jacobson_2011_Metallicity_gradient}.

We also compared our membership with the C18 membership for our GOOD clusters. Considering stars with APOGEE observations, C18 has about 5\% more members for each cluster than we do, but these stars typically have RVs inconsistent with the peak of the cluster. For a few clusters, we find additional members (about 4\% of the total) that are not present in C18. These stars do not have measured {\it Gaia}-DR2 parallaxes, and we believe that this is the reason they have been excluded from C18. However, these additional members we find do have measured RA17, StarHorse, and astroNN distances (\S\ref{sec:data_distances}) that are generally similar to the distances of the members common to both membership catalogs. We repeated the analysis in \S\ref{sec:results} using only the common members and found similar results and interpretations.

\begin{table*}[]
\centering
\begin{tabular}{l|c|c|c|c|c|c}
\multirow{2}{*}{APOGEE\_ID} & \multirow{2}{*}{CLUSTER} & RA & DEC & \multirow{2}{*}{NO\_SIGMAS\_RV} & \multirow{2}{*}{NO\_SIGMA\_PM} & DIST\_CENTER \\
 & & deg & deg & & & arcmin \\
\hline
\hline
2M00000068+5710233 & NGC 7789 &	0.0029 & 57.1732 & 37.79 & 24.49 & 33.99\\
2M00001199+6114138 & NGC 7790 &	0.0500 & 61.2372 & 12.24 & 3.93 & 13.36\\
2M00001328+5725563 & NGC 7789 &	0.0554 & 57.4323 & 8.70 & 5.09 & 47.91\\
2M00002012+5612368 & NGC 7789 &	0.0839 & 56.2102 & 33.14 & 2.58 & 39.47\\
2M00002853+6119307 & NGC 7790 &	0.1189 & 61.3252 & 5.35 & 1.33 & 16.94\\
\end{tabular}
\caption{Sample table of cluster members selected using our membership selection~\S\ref{sec:membership}.}
\label{tab:cluster_members_sample_table}
\tablecomments{Table~\ref{tab:cluster_members_sample_table} is published in its entirety in the machine-readable format. A portion is shown here for guidance regarding its form and content.}

\end{table*}

\begin{table*}[]
\centering
\begin{tabular}{l|l}
Column  &  Description \\
\hline
\hline
NAME & Name of cluster \\
CENTER\_RA & Central right ascension$^1$ [deg] \\
CENTER\_DEC & Central declination$^1$ [deg] \\
RADIUS & Adopted cluster radius$^1$ [arcmin] \\
DISTANCE\_APOGEE & Median RA17 spectrophotometric distance of cluster members [kpc] (\S\ref{sec:data_distances})\\
DISTANCE\_DISP\_APOGEE & Dispersion of spectrophotometric distance [kpc] \\
LOG\_AGE & Log(age) of cluster$^1$ [dex]\\
LOG\_AGE\_ERR & Uncertainty in log(age) of cluster$^1$ [dex]\\
M\_H & Mean [M/H] of cluster members [dex] (\S\ref{sec:results_no_trend})\\
M\_H\_DISP & Standard deviation of [M/H] of cluster members [dex]\\
RV\_FIT\_MEAN & Mean of best-fit Gaussian to RVs [km/s] (\S\ref{sec:RVs})\\
RV\_FIT\_STD & Standard deviation of best-fit Gaussian to RVs [km/s] (\S\ref{sec:RVs})\\
RV\_FIT\_AMP & Amplitude of best-fit Gaussian to RVs (\S\ref{sec:RVs})\\
PM\_FIT\_RA\_MEAN & Mean $\mu_\alpha$ of best-fit Gaussian to $\mu_\alpha \times \mu_\delta$ [mas/year] (\S\ref{sec:PMs})\\
PM\_FIT\_DEC\_MEAN & Mean $\mu_\delta$ of best-fit Gaussian to $\mu_\alpha \times \mu_\delta$ [mas/year] (\S\ref{sec:PMs})\\
PM\_FIT\_RA\_STD & Standard deviation in RA of best-fit Gaussian to $\mu_\alpha \times \mu_\delta$ [mas/year] (\S\ref{sec:PMs})\\
PM\_FIT\_DEC\_STD & Standard deviation in DEC of best-fit Gaussian to $\mu_\alpha \times \mu_\delta$ [mas/year] (\S\ref{sec:PMs})\\
PM\_FIT\_AMP & Amplitude of best-fit Gaussian to $\mu_\alpha \times \mu_\delta$ (\S\ref{sec:PMs})\\
PM\_FIT\_THETA & Rotation angle [rad] (\S\ref{sec:PMs})\\
PM\_FIT\_BACKGROUND\_SCALE & Scale of the annular PM distribution (\S\ref{sec:PMs})\\
AMP\_RESIDUAL\_RV & Amplitude over residual for RV (\S\ref{sec:RVs})\\
NUM\_MEMBERS & Number of selected cluster members\\
FLAG & Validation flag for the cluster (\S\ref{sec:validation})\\
MG\_FE & Mean [Mg/Fe] of cluster members [dex]\\
MG\_FE\_DISP & Standard deviation of [Mg/Fe] of cluster members [dex]\\
Z & Vertical distance from the Milky Way disc [kpc] (\S\ref{sec:results_no_trend})\\
R\_GC & Galactocentric distance of cluster [kpc] (\S\ref{sec:results_no_trend}) \\
R\_GC\_ERR & Uncertainty in R\_GC [kpc]\\
SPACE\_VEL\_DISP & 3D velocity dispersion [kpc] (\S\ref{sec:velocity_dispersion}) \\
SPACE\_VEL\_DISP\_ERR & Uncertainty in 3D velocity dispersion\\
\hline
\end{tabular}
\caption{Columns from the table of catalog clusters.
\vspace{0pt}
$^1$From K13}
\label{tab:catalog_example_clusters}
\tablecomments{Table~\ref{tab:catalog_example_clusters} is published in its entirety in the machine-readable format. A portion is shown here for guidance regarding its form and content.}
\end{table*}

\begin{table*}[]
\centering

\begin{tabular}{c|c|c|c|c|c|c|c}
{\tiny \multirow{3}{*}{NAME}} & {\tiny DISTANCE\_APOGEE$\pm$}  & {\tiny RV\_FIT\_MEAN$\pm$} & {\tiny PM\_FIT\_RA\_MEAN$\pm$} & {\tiny PM\_FIT\_DEC\_MEAN$\pm$} & {\tiny \multirow{3}{*}{NUM\_MEMBERS}} & {\tiny M\_H$\pm$} & {\tiny MG\_FE$\pm$} \\

& {\tiny DISTANCE\_DISP\_APOGEE} & {\tiny RV\_FIT\_STD} & {\tiny PM\_FIT\_RA\_STD} & {\tiny PM\_FIT\_DEC\_STD} & & {\tiny M\_H\_DISP} & {\tiny MG\_FE\_DISP} \\

& kpc & km~s$^{-1}$ & mas~yr$^{-1}$ & mas~yr$^{-1}$ & & dex & dex \\
\hline
\hline
NGC 1245 & 3.19$\pm{0.19}$ & -29.18$\pm{0.79}$ & 0.55$\pm{0.57}$ & -1.67$\pm{0.49}$ & 26 & -0.080$\pm{0.025}$ & -0.028$\pm{0.024}$ \\

NGC 188 & 1.85$\pm{0.18}$& -41.96$\pm{0.33}$ & -2.32$\pm{0.54}$ & -0.94$\pm{0.52}$ & 29  & 0.100$\pm{0.029}$ & 0.033$\pm{0.026}$ \\

NGC 2204 & 3.69$\pm{1.19}$ & 92.09$\pm{1.01}$ & -0.54$\pm{0.55}$ & 1.96$\pm{0.51}$ & 27 & -0.282$\pm{0.096}$ & 0.014$\pm{0.049}$ \\

NGC 2420 & 2.19$\pm{0.43}$ & 74.22$\pm{0.93}$ & -1.15$\pm{0.50}$ & -2.16$\pm{0.59}$ & 18 & -0.201$\pm{0.067}$ & 0.004$\pm{0.027}$ \\

NGC 2682 & 0.75$\pm{0.13}$ & 34.05$\pm{0.66}$ & -10.98$\pm{0.55}$ & -2.95$\pm{0.56}$ & 381 & -0.007$\pm{0.058}$ & 0.004$\pm{0.034}$ \\

NGC 6705 & 0.93$\pm{0.61}$& 35.51$\pm{1.65}$ & -2.31$\pm{1.29}$ & -5.05$\pm{0.79}$ & 15  & 0.172$\pm{0.056}$ & -0.058$\pm{0.036}$ \\

NGC 6791 & 4.65$\pm{0.95}$& -47.05$\pm{1.39}$ & -0.42$\pm{0.52}$ & -2.27$\pm{0.51}$ & 59  & 0.346$\pm{0.050}$ & 0.099$\pm{0.034}$ \\

NGC 6819 & 2.37$\pm{0.61}$& 2.74$\pm{1.18}$ & -2.93$\pm{0.54}$ & -3.88$\pm{0.57}$ & 48  & 0.057$\pm{0.099}$ & -0.008$\pm{0.022}$ \\

NGC 7789 & 1.97$\pm{0.76}$ & -54.77$\pm{1.12}$ & -0.91$\pm{0.51}$ & -1.95$\pm{0.53}$ & 25 & -0.018$\pm{0.085}$ & -0.017$\pm{0.033}$ \\

VDB 131 & 2.39$\pm{0.53}$& -31.89$\pm{1.68}$ & -6.00$\pm{0.67}$ & 0.19$\pm{0.72}$ & 13  & 0.152$\pm{0.071}$ & -0.038$\pm{0.033}$ \\
\hline
\end{tabular}
\caption{A subsample of columns from Table~\ref{tab:catalog_example_clusters} for the ten clusters used in \S\ref{sec:results}. Shown are the mean cluster heliocentric distances, metallicities, and [Mg/Fe] abundances, and the means of the Gaussian fits for each kinematic dimension, along with their standard deviations. `VDB 131' is short for VDBERGH-HAGEN 131.}
\label{tab:subsample_clusters}
\end{table*}

\section{Quantifying Cluster Homogeneity}
\label{sec:homogeneity}

To study the chemical homogeneity of the kinematically identified cluster members in \S\ref{sec:membership}, we need a robust homogeneity metric that takes into account the members' non-uniform abundance uncertainties (\S\ref{sec:uncertainties}).  Previous studies of intrinsic abundance scatter adopted a variety of metrics, 
such as the root-mean square of the abundances \citep[][]{De_Silva_2007_Collinder_261_homogeneity}
and a $\chi^{2}$-like measurement of the distance between pairs of
stars in an $N$-dimensional chemical space \citep[][]{Ness_2018_doppelgangers}.

We adopt a Maximum Likelihood Estimator (MLE) approach to determining the intrinsic abundance scatter of a group of stars, similar to the MLE in \citet{Kovalev_2019_OC_scatter}. This choice is based on the speed and simplicity of the method, combined with its consistency with other tested metrics (see below). 

\begin{figure}[h]%
	\centering
    \includegraphics[scale=0.7]{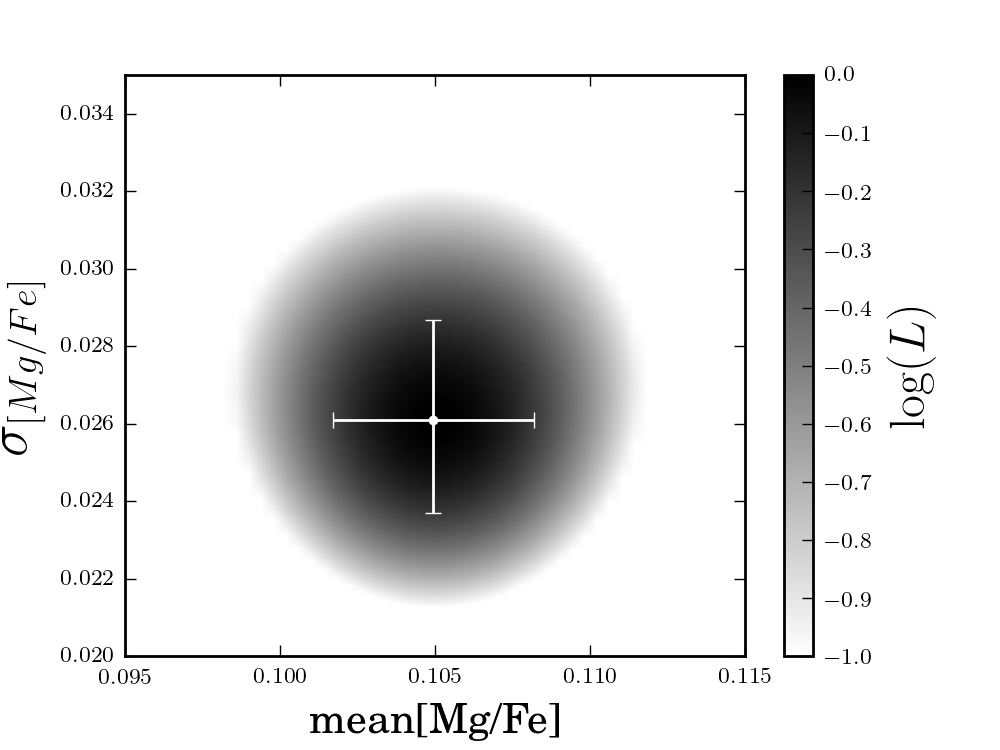}
    \caption{
    Example of the $\mu_{\rm[X/Fe]}-\sigma_{\rm[X/Fe]}$ plane of the likelihood function (Eq.~\ref{eqn:MLE}) used to determine the intrinsic scatter ($\sigma_{\rm[X/Fe]}$) for [Mg/Fe] in NGC 6791.
    }
    \label{fig:MLE}
\end{figure}

Given a distribution of abundances [X/Fe] with their corresponding uncertainties, we can estimate the likelihood that these values were drawn from a Gaussian distribution centered at $\mu_{\rm[X/Fe]}$ with a standard deviation 
of $\sigma_{\rm[X/Fe]}$ using:
\begin{equation}
\label{eqn:MLE}
	L \equiv \prod_{i=1}^{n} \frac{1}{\sqrt{2 \pi}(\sigma_{\rm[X/Fe]}^2+e_i^2)^{1/2}} \exp\Bigg(\frac{-(x_i-\mu_{\rm[X/Fe]})^2}{2(\sigma_{\rm[X/Fe]}^2+e_i^2)}\Bigg),
\end{equation}
where $x_i$ is the chemical abundance of a particular element for a cluster member and $e_i$ is
the corresponding abundance uncertainty. By finding where the maximum of this function
lies in the $\mu_{\rm[X/Fe]}-\sigma_{\rm[X/Fe]}$ plane shown in Figure~\ref{fig:MLE}, we can estimate the parameters of the Gaussian
distribution from which these data points are drawn. Here, we are most interested in the value
of $\sigma_{\rm[X/Fe]}$, since it represents the intrinsic scatter of the abundances within the cluster. We estimate the asymmetric uncertainty in the value of $\sigma_{\rm[X/Fe]}$ using the distribution of the likelihood function along the $\sigma_{\rm[X/Fe]}$ axis. We take the first and third quartile ranges of this distribution as the lower and upper uncertainty limits on $\sigma_{\rm[X/Fe]}$.

We verified that this method can recover an input $\sigma_{\rm[X/Fe]}$ value from mock abundances that have been perturbed by uncertainties assigned from stars in several of our clusters for a given element. During this test, we noticed the existence of a bias for the MLE estimator with respect to the number of stars in each cluster -- specifically, clusters with fewer stars ($N < 15$) were systematically estimated to have lower scatter than the actual value. We resolved this issue by fitting this bias with an exponential function and scaling the derived MLE scatter based on the number of members in each cluster. 

We looked at the distribution of individual stellar APOGEE RVs versus abundances in clusters to ensure that the calculated value of $\sigma_{\rm[X/Fe]}$ was not being driven by outlier stars in each dimension.
We ensured that there was no evident trend between intrinsic scatter and dispersion of $T_{\rm eff}$ or logg for the clusters that we study. We verified for multiple clusters that the elemental abundance distributions followed a Gaussian distribution since this is an assumption intrinsic to the MLE method. We also studied how the $\sigma_{\rm[X/Fe]}$ changed for stars that belonged to different evolutionary stages within the same cluster. 

As a consistency check, we compared the MLE-based $\sigma_{\rm[X/Fe]}$ to that estimated from other metrics.  One other metric we considered compares the cumulative distribution of pairwise distances in $N$-dimensional chemical space of simulated abundances with real cluster data.
This metric is more computationally expensive than the MLE method but produces results that are entirely consistent.

\section{RESULTS}
\label{sec:results}

\subsection{Final selection of elements and cluster members}
\label{sec:results_elements_members}

We select elements for our analysis from the full set available in APOGEE using a variety of criteria.
Some elements are known to have issues with accurate abundance determinations with ASPCAP, at least in certain ranges of stellar parameter relevant to our stars \citep[e.g., S, K, Na, and Ti;][]{Hawkins_2016_abundance_uncertainties}, and we discard these. We also remove C, N, and O from further analysis since the abundances of these elements are affected by different stages of dredge-up over the course of the evolution of the star.

We use the re-derived uncertainties (\S\ref{sec:uncertainties}) to calculate and compare the chemical homogeneity of OCs to that of groups of field stars selected to match the clusters in spatial extent and stellar parameters. A few elements (e.g., P, V, and Ce) for which field stars sample show a lower scatter in abundance are removed from further analysis, since field stars chosen in this way are expected to have higher scatter in abundances than OCs.
Based on these quality cuts, we only use [Mg/Fe], [Al/Fe], [Si/Fe], [Ca/Fe], [Cr/Fe], [Mn/Fe], [Ni/Fe], and [Fe/H] for the analyses in this section. 

The intrinsic chemical scatter within clusters ($\sigma_{\rm[X/Fe]}$) used in this section are derived from GOOD clusters (\S\ref{sec:validation}) with at least nine members (\S\ref{sec:probabilities}) that are within 1$R_{\rm cluster}$ from the cluster center and that meet the following APOGEE bitmask\footnote{\url{https://www.sdss.org/dr16/algorithms/bitmasks/}} criteria:

\begin{itemize} \itemsep -2pt
    \item BRIGHT\_NEIGHBOR and VERY\_BRIGHT\_NEIGHBOR == 0 (STARFLAG bits 2 and 3)
    \item SUSPECT\_BROAD\_LINES==0 (STARFLAG bit 17)
    \item METALS\_BAD==0 (ASPCAPFLAG bit 19)
    \item ALPHAFE\_BAD==0 (ASPCAPFLAG bit 20)
    \item STAR\_BAD==0 (ASPCAPFLAG bit 23)
\end{itemize}

We further restrict our sample to giant stars (using the ASPCAP\_CLASS designation and a limit of $\log{g}<3$) with S/N$>$50. 
This $\log{g}$ limit is implemented to remove stars whose abundances could potentially be affected by atomic diffusion~\citep{Souto_2019_M67_atomic_diffusion,Semenova_2020_atomic_diffusion}.
Finally, we remove stars that lie in ranges of $T_{\rm eff}$, [M/H], and SNR in which the distribution of observed visit-to-visit abundance variations (\S\ref{sec:uncertainties_calculate}) is non-Gaussian. We find that in a small number ($\sim$3\%) of the stellar parameter bins used to derive abundance uncertainties, a significant fraction of the stellar visit pairs result in abundance differences $>$0.5~dex.  Cluster members that fall in these bins are removed from measurements of abundance scatter of that particular element, since the computed abundance uncertainty may not reflect the true deviation from the correct answer.

These limits result in ten GOOD OCs with sufficient members in all elements for further analysis.
We explored several combinations of these limits (e.g., the minimum number of stars required in each OC, minimum SNR condition), all for which the final results and interpretations remain the same as described below. The membership plots for these selected clusters are included in Appendix \S\ref{sec:app_membership}. Note that most of the outliers in the MDFs of the clusters (panel f, shown in purple) fail to pass the quality cuts mentioned above and so are not included in the analysis.

\subsection{Cluster distances}
We use stellar distances to compute median cluster heliocentric distances, which are used to calculate Galactocentric distance ($R_{\rm GC}$), height above the midplane ($|Z|$) (\S\ref{sec:results_no_trend}), and space space velocity dispersion (\S\ref{sec:velocity_dispersion}).  
We find extremely good agreement in these median distances using four distance catalogs: StarHorse, astroNN, RA17 (\S\ref{sec:data_distances}) and distances calculated using the \textit{Gaia}-DR2 parallax of the cluster members. 

The two exceptions are VDBERGH-HAGEN~131 and NGC 6705, where the four catalogs give median distances that vary by a factor of $\sim$2. VDBERGH-HAGEN~131 stands out as being the most heavily reddened ($E(J-K_s) \sim 0.6$), and the most differentially reddened ($\sigma_{E(J-K_s)} \sim 0.12$), as evident in its CMD in Appendix~\ref{sec:app_membership}. We observed that the CMD for VDBERGH-HAGEN~131 dereddened with the RA17 reddening estimates produces a tighter locus than with StarHorse and astroNN. Additionally, for both VDBERGH-HAGEN~131 and NGC 6705, the RA17 distances for the members have a slightly smaller dispersion, compared to the StarHorse, astroNN and \textit{Gaia}-DR2 parallax-based distance values.

So we adopt the RA17 distances for these two clusters, and for consistency for all of the clusters.  We emphasize that the results described below are independent of the catalog used to calculate the distance.

\subsection{Abundance scatter in clusters}
\label{abundance_scatter}
We calculate the abundance scatter in ten OCs for 8 elements (Mg, Al, Si, Ca, Fe, Si, Mn, and Ni) using the method discussed in \S\ref{sec:homogeneity}. We measure non-zero intrinsic scatter ($\sigma_{\rm[X/Fe]}$) in most cases. 
From Table~\ref{tab:cluster_scatters}, we see that all clusters except NGC 2204, NGC 6791, and VDBERGH-HAGEN 131 have $\sigma_{\rm[Fe/H]}$ very close to previously determined limits for scatter in [Fe/H]~\citep[a range of 0.02 -- 0.04 dex;][]{Bovy_2016_OC_homogeneity,De_Silva_2007_Collinder_261_homogeneity,Kovalev_2019_OC_scatter}. Two of these three have $\sigma_{\rm[Fe/H]}$ less than 0.05 dex, with the exception of VDBERGH-HAGEN 131, which is a lesser studied cluster with no previous abundance determinations or abundance scatter studies performed.

VDBERGH-HAGEN 131 also exceeds the limit (0.03 dex) predicted by \citet{Bovy_2016_OC_homogeneity} for $\sigma_{\rm[Al/Fe]}$. Although chemical abundances have been determined for some red giants in NGC 2204~\citep[e.g.,][]{Jacobson_2011_NGC_2204,Carlberg_2016_NGC_2204}, there have been no studies focused on its chemical homogeneity. We find $\sigma_{\rm[Mg/Fe]}$, $\sigma_{\rm[Al/Fe]}$, and $\sigma_{\rm[Si/Fe]}$ in NGC 2204 to be higher than average literature limits ($\sim$0.03 dex) for other OCs.

However, the most interesting case we observe is NGC 6791, a high-metallicity OC whose chemistry has been well studied \citep[e.g.,][]{Cunha_2015_NGC_6791}. We measure a value of $\sigma_{\rm[Mn/Fe]}$ for NGC 6791 that is very high compared to the $\sigma_{\rm[Mn/Fe]}$ values for the rest of our clusters. The $\sigma_{\rm[X/Fe]}$ values for the other elements in NGC 6791 fall within the limits quoted by \citet{Bovy_2016_OC_homogeneity}, except for $\sigma_{\rm[Al/Fe]}$ (limit $\sim$0.03 dex). \citet{Donor_2020_occam3} also report a particularly high uncertainty of 0.13 dex in their mean [Mn/Fe] for this cluster, where their uncertainty is defined as the 1-sigma scatter in cluster [Mn/Fe] abundances in APOGEE. We have verified that this atypically high measurement of $\sigma_{\rm[Mn/Fe]}$ is not a result of non-members with discrepant [M/H] measurements that may have been selected as members (e.g., $\sigma_{\rm[Fe/H]}<0.05$ for this cluster, which is highly unlikely if contamination were large). 
We have verified that Mn lines for NGC 6791 members can be reliably measured over a range of $T_{\rm eff}$ at high [M/H]. We also find no systematic increase in random uncertainties at higher [M/H] nor any systematic shift in [Mn/Fe] abundances with $T_{\rm eff}$.

We compared abundance scatter between elements that are observed to have a high abundance variations in GCs and those that are not.
Of the elements that are included in our study, Mg, Al and in few cases Si are those that have confirmed observations of significant abundance scatter and anti-correlations in GCs~\citep{Gratton_2020_GC_review}. 
As described above, the abundance scatter in Mg, Al, and Si for NGC 2204 stands out above the literature limits for OCs.
However, we do not observe a selectively higher abundance scatter in these elements for any of our other OCs. 
\begin{table*}[]
\centering
\begin{tabular}{c|c|c|c|c|c|c|c|c|c}
\multirow{2}{*}{Cluster}  &  $\sigma_{\rm[Fe/H]}$ & $\sigma_{\rm[Mg/Fe]}$ & $\sigma_{\rm[Al/Fe]}$ & $\sigma_{\rm[Si/Fe]}$ & $\sigma_{\rm[Ca/Fe]}$ & $\sigma_{\rm[Cr/Fe]}$ & $\sigma_{\rm[Mn/Fe]}$ & $\sigma_{\rm[Ni/Fe]}$ & $\sigma_{\rm tot}$ \\
 & dex & dex & dex & dex & dex & dex & dex & dex & km~s$^{-1}$ \\
\hline
\hline
NGC 1245 & \tiny 0.0211$\pm^{0.003}_{0.002}$ & \tiny 0.0246$\pm^{0.003}_{0.003}$ & \tiny 0.0233$\pm^{0.005}_{0.005}$ & \tiny 0.0154$\pm^{0.003}_{0.002}$ & \tiny 0.0252$\pm^{0.004}_{0.003}$ & \tiny 0.0582$\pm^{0.009}_{0.008}$ & \tiny 0.0000$\pm^{0.004}_{0.003}$ & \tiny 0.0127$\pm^{0.002}_{0.002}$ & \tiny 5.41$\pm1.32$\\

NGC 188 & \tiny 0.0219$\pm^{0.006}_{0.004}$ & \tiny 0.0143$\pm^{0.005}_{0.004}$ & \tiny 0.0341$\pm^{0.010}_{0.008}$ & \tiny 0.0000$\pm^{0.003}_{0.002}$ & \tiny 0.0148$\pm^{0.004}_{0.004}$ & \tiny 0.0135$\pm^{0.009}_{0.008}$ & \tiny 0.0240$\pm^{0.008}_{0.006}$ & \tiny 0.0074$\pm^{0.004}_{0.003}$ & \tiny 3.10$\pm0.77$\\

NGC 2204 & \tiny 0.0422$\pm^{0.007}_{0.006}$ & \tiny 0.0414$\pm^{0.007}_{0.006}$ & \tiny 0.0511$\pm^{0.011}_{0.009}$ & \tiny 0.0443$\pm^{0.008}_{0.007}$ & \tiny 0.0224$\pm^{0.006}_{0.005}$ & \tiny 0.0000$\pm^{0.009}_{0.006}$ & \tiny 0.0134$\pm^{0.007}_{0.006}$ & \tiny 0.0068$\pm^{0.004}_{0.004}$ & \tiny 6.49$\pm1.58$\\

NGC 2420 & \tiny 0.0314$\pm^{0.007}_{0.006}$ & \tiny 0.0160$\pm^{0.005}_{0.004}$ & \tiny 0.0210$\pm^{0.007}_{0.006}$ & \tiny 0.0000$\pm^{0.003}_{0.002}$ & \tiny 0.0202$\pm^{0.006}_{0.005}$ & \tiny 0.0430$\pm^{0.014}_{0.011}$ & \tiny 0.0157$\pm^{0.006}_{0.005}$ & \tiny 0.0000$\pm^{0.003}_{0.002}$ & \tiny 4.18$\pm0.99$\\

NGC 2682 & \tiny 0.0263$\pm^{0.004}_{0.003}$ & \tiny 0.0148$\pm^{0.003}_{0.003}$ & \tiny 0.0132$\pm^{0.005}_{0.005}$ & \tiny 0.0153$\pm^{0.003}_{0.003}$ & \tiny 0.0000$\pm^{0.003}_{0.002}$ & \tiny 0.0728$\pm^{0.012}_{0.010}$ & \tiny 0.0134$\pm^{0.004}_{0.004}$ & \tiny 0.0028$\pm^{0.002}_{0.002}$ & \tiny 1.68$\pm0.35$\\

NGC 6705 & \tiny 0.0359$\pm^{0.008}_{0.006}$ & \tiny 0.0124$\pm^{0.004}_{0.003}$ & \tiny 0.0275$\pm^{0.008}_{0.006}$ & \tiny 0.0112$\pm^{0.003}_{0.003}$ & \tiny 0.0161$\pm^{0.004}_{0.003}$ & \tiny 0.0326$\pm^{0.009}_{0.007}$ & \tiny 0.0096$\pm^{0.005}_{0.005}$ & \tiny 0.0072$\pm^{0.004}_{0.004}$ & \tiny 3.50$\pm0.68$\\

NGC 6791 & \tiny 0.0491$\pm^{0.004}_{0.004}$ & \tiny 0.0268$\pm^{0.003}_{0.003}$ & \tiny 0.0709$\pm^{0.007}_{0.006}$ & \tiny 0.0205$\pm^{0.003}_{0.002}$ & \tiny 0.0263$\pm^{0.003}_{0.003}$ & \tiny 0.0693$\pm^{0.008}_{0.007}$ & \tiny 0.1146$\pm^{0.011}_{0.010}$ & \tiny 0.0285$\pm^{0.003}_{0.003}$ & \tiny 7.59$\pm1.83$\\

NGC 6819 & \tiny 0.0343$\pm^{0.003}_{0.003}$ & \tiny 0.0081$\pm^{0.002}_{0.002}$ & \tiny 0.0329$\pm^{0.004}_{0.003}$ & \tiny 0.0211$\pm^{0.002}_{0.002}$ & \tiny 0.0174$\pm^{0.002}_{0.002}$ & \tiny 0.0246$\pm^{0.005}_{0.005}$ & \tiny 0.0252$\pm^{0.003}_{0.003}$ & \tiny 0.0100$\pm^{0.002}_{0.003}$ & \tiny 4.24$\pm0.98$\\

NGC 7789 & \tiny 0.0318$\pm^{0.005}_{0.004}$ & \tiny 0.0043$\pm^{0.003}_{0.003}$ & \tiny 0.0300$\pm^{0.006}_{0.005}$ & \tiny 0.0102$\pm^{0.003}_{0.003}$ & \tiny 0.0128$\pm^{0.003}_{0.003}$ & \tiny 0.0000$\pm^{0.007}_{0.006}$ & \tiny 0.0245$\pm^{0.004}_{0.004}$ & \tiny 0.0000$\pm^{0.003}_{0.002}$ & \tiny 3.23$\pm0.71$\\

VDB 131 & \tiny 0.0771$\pm^{0.018}_{0.014}$ & \tiny 0.0205$\pm^{0.009}_{0.007}$ & \tiny 0.0576$\pm^{0.002}_{0.012}$ & \tiny 0.0123$\pm^{0.005}_{0.004}$ & \tiny 0.0183$\pm^{0.006}_{0.005}$ & \tiny 0.0357$\pm^{0.013}_{0.011}$ & \tiny 0.0331$\pm^{0.010}_{0.007}$ & \tiny 0.0161$\pm^{0.007}_{0.005}$ & \tiny 5.81$\pm1.33$\\
\hline
\end{tabular}
\caption{Intrinsic abundance scatter (\S\ref{sec:homogeneity}) and space velocity dispersion ($\sigma_{\rm tot}$; Equation~\ref{eqn:space_velocity_disp}) for the OCs analyzed in \S\ref{sec:results}. `VDB 131' is short for VDBERGH-HAGEN 131.}
\label{tab:cluster_scatters}
\end{table*}

\subsection{Galactic position, age, and  metallicity}
\label{sec:results_no_trend}

\begin{figure*}
\begin{center}
\includegraphics[angle=0, clip, width=\textwidth]{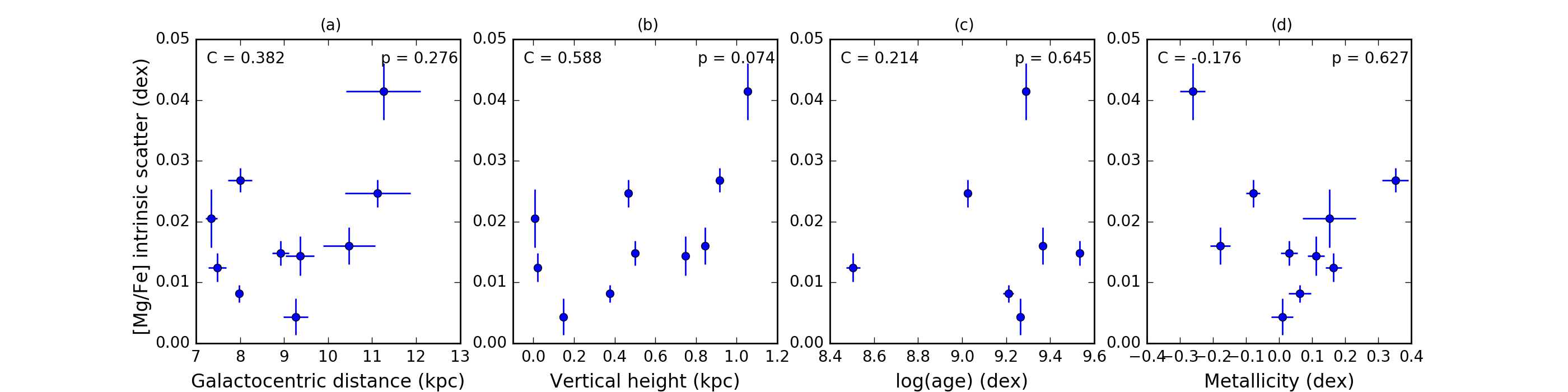}
\caption{
Dependence of cluster [Mg/Fe] homogeneity on Galactocentric distance, vertical height, log(age), and metallicity of the cluster. The Spearman correlation coefficient (C) and corresponding p-value (p) are shown for each property.
} 
\label{fig:no_trend_properties}
\end{center}
\end{figure*}  

We find that cluster abundance scatter is uncorrelated with Galactocentric distance and vertical height from the plane of the Milky Way for all the elements we consider. Figure~\ref{fig:no_trend_properties}a--b shows examples of the trend of cluster scatter in [Mg/Fe] with respect to Galactocentric distance and vertical height, respectively. 

For ages we use values from the K13 catalog that have reported uncertainties in their age measurements (seven out of the ten clusters). We find that cluster scatter is uncorrelated with cluster age. An example plot is shown for [Mg/Fe] in Figure \ref{fig:no_trend_properties}c. We calculate the mean metallicity ([M/H]) of each cluster, using its APOGEE members, and find that metallicity is uncorrelated with cluster scatter. An example plot is shown for [Mg/Fe] in Figure \ref{fig:no_trend_properties}d. 

In Figure~\ref{fig:no_trend_properties}b, although we see a relatively higher correlation coefficient compared to the rest of the subplots, we do not believe that this shows the presence of a significant correlation since this trend is not evident in any other element ([X/Fe] or [Fe/H]) that we consider. 
We also looked for correlations between chemical scatter and these Galactic/cluster properties in selected subgroups, such as thin and thick disk clusters, but did not find anything of significance.

We examined whether cluster scatter was correlated with physical cluster size, which we calculated using the angular cluster radius from K13 and the median stellar distance, and found no relationship.
However, we note that we consider these size values to be highly uncertain, since they depend on the choice of angular radius definition and in at least some cases, clearly do not match the kinematically-clumped stars at that location.

\subsection{Velocity dispersion (cluster mass)}
\label{sec:velocity_dispersion}

\subsubsection{Correlation with velocity dispersion}
\label{sec:velocity_dispersion_correlation}

We calculate the 3D velocity dispersion of a cluster, a proxy for cluster mass, from its RV and PM dispersions and heliocentric distance using the following equation:
\begin{equation}
\label{eqn:space_velocity_disp}
    \sigma_{\rm tot} = \sqrt{\sigma_{RV}^2+ (\sigma_{\mu_{\alpha}}^2+\sigma_{\mu_{\delta}}^2)d_{helio}^2}
\end{equation}{}
where $\sigma_{\rm tot}$ is the space velocity dispersion, corrected by the uncertainties as described in \S\ref{sec:homogeneity}; $\sigma_{RV}, \sigma_{\mu_{\alpha}},$ and $\sigma_{\mu_{\delta}}$ are the dispersions in the cluster for each kinematic dimension; and $d_{helio}$ is the heliocentric distance, assigned as the median of the stellar member distances from RA17 (\S\ref{sec:data_distances}). We observe a strong correlation between the calculated RV and PM velocity dispersions, which ensures that $\sigma_{\rm tot}$ is not being driven by any one dimension alone.

We find that the cluster chemical scatter is positively correlated with the space velocity dispersion of the cluster at relatively high levels of significance ($p < 0.019$) for [Mg/Fe] and [Ca/Fe]; possibly positively correlated at a low level of significance ($0.038 < p < 0.059$) for [Ni/Fe], [Si/Fe], [Al/Fe], and [Fe/H]; and uncorrelated ($p > 0.38$) for [Cr/Fe] and [Mn/Fe]. Figure~\ref{fig:abundance_scatter_vs_dispersion} shows the intrinsic scatter in [Fe/H] and rest of the abundances as a function of space velocity dispersion, along with the associated Spearman correlation coefficients (C) and $p$-values.

To understand why only certain elements show this trend between intrinsic scatter and $\sigma_{\rm tot}$, we look for natural ways to group elements based on their properties.
For example, we notice that this trend is not exclusive to the $\alpha$-elements that we study. Although intrinsic scatter is positively correlated with $\sigma_{\rm tot}$ in Mg, Ca, and Si (albeit at low significance), we observe a similar trend in an odd-Z element (Al) and an iron-peak element (Ni) at lower significance. In \S\ref{abundance_scatter} we discuss how the abundance scatters we observe behave for elements that show abundance variations and anti-correlations in GCs (e.g., Mg, Al, and Si). Again, this trend is not restricted to these three elements. However, we find that if we group elements based on their dominant nucleosynthetic process of production, we see a distinction between elements that are produced predominantly by core-collapse supernovae (CCSNe) versus Type~Ia SNe.

In order to visualize the differences in strengths of correlation in groups of elements and to explore the roles that different enrichment events may have played in them, we use the empirically determined fractional contribution of CCSNe ($f_{cc}$) for each element from \citet{Weinberg_2019_chemical_cartography}. 
Figure~\ref{fig:abundance_scatter_vs_fcc} shows the Spearman correlation coefficients from Figure~\ref{fig:abundance_scatter_vs_dispersion} against $f_{cc}$, with the points colored by the $p$-value of their correlation.
Here $f_{cc}$ represents the fraction of each element contributed by CCSNe at a given metallicity [Mg/H], assuming that these elements are produced exclusively by Type~Ia SNe and CCSNe.  We calculate $f_{cc}$ from Equation~11 in \citet{Weinberg_2019_chemical_cartography}, using the median [Mg/H] value for each of our clusters.

From Figure~\ref{fig:abundance_scatter_vs_fcc}, in addition to metallicity (traced by [Fe/H]), the elements that show a correlation between intrinsic abundance scatter and cluster velocity dispersion (with C $>$ 0.6) at both higher (with $p < 0.019$) and lower ($0.038 < p < 0.059$) levels of significance are the ones that are produced mostly by CCSNe.

\subsubsection{Caveats of the significance of the correlations}
\label{sec:velocity_dispersion_caveats}

While these correlations between abundance scatter and space velocity dispersion are interesting, we note that the statistical analysis is done using only ten clusters and the elements that we list as correlated have varying levels of significance ($p$-values). Here we explore the caveats associated with these correlations. One effect of the low sample size is seen when we randomly remove any one cluster from the analysis using a jackknife resampling. For certain elements, the correlation becomes insignificant if we remove a specific cluster (e.g., removing NGC~2204 in the case of Si or removing NGC~6791 in the case of Ni) from the analysis. However, we found that no one cluster is responsible for systematically reducing the significance across the set of all the elements. Since we arrive at this cluster sample by preferentially selecting clusters with high quality membership and reliable abundance uncertainties (\S\ref{sec:results_elements_members}), we have no a priori reason to drop any particular cluster from any particular elemental trend.

We explored the effect of using a different correlation metric (Kendall's tau) on the strength and significance of the correlations seen in Figure~\ref{fig:abundance_scatter_vs_dispersion}. We do not use the Pearson metric in this comparison since it is susceptible to outliers and assumes linear relationships. The Spearman and Kendall correlation metrics are highly correlated themselves, though their magnitude is not equal, and they more robust to outliers since they use the ranks of the variables rather than the actual values. We show the Spearman metric in Figures~\ref{fig:abundance_scatter_vs_dispersion} and \ref{fig:abundance_scatter_vs_fcc} since it is more commonly used, although we note that Kendall may be more accurate for small sample sizes. The significance of the correlations for Al, Si, Fe, and Ni depends somewhat on the choice of the metric, with Kendall's tau associated with $p$-values up to 0.07, which reinforces our classification of these elements as possible correlations. However, Mg and Ca scatters are significantly correlated with velocity dispersion irrespective of the metric used, while Mn and Cr remain uncorrelated.

\subsubsection{Potential Implications}
\label{sec:velocity_dispersion_implications}

\begin{figure*}
\begin{center}
   
   \includegraphics[angle = 0, clip, scale = 0.55]{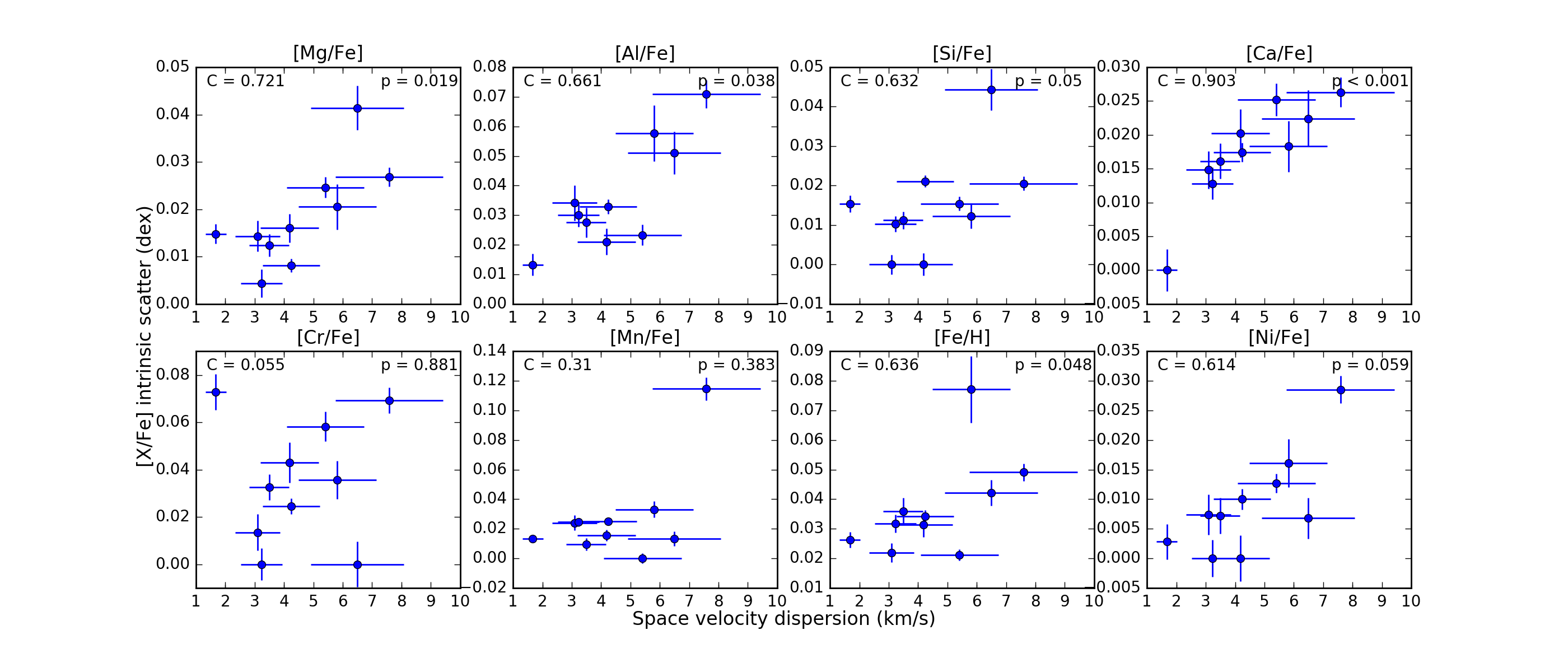}
   \caption{Dependence of cluster [X/Fe] homogeneity on space velocity dispersion. The Spearman correlation coefficients (C) and corresponding p-values (p) are shown for each element.}
   \label{fig:abundance_scatter_vs_dispersion}

\end{center}

\end{figure*}

As discussed in \S\ref{sec:velocity_dispersion_correlation}, it is interesting that the elements that show some level of correlation with space velocity dispersion (Figure~\ref{fig:abundance_scatter_vs_dispersion}) are those that are predominantly produced by CCSNe.
However, due to the small sample size, both in clusters and elements, and other caveats discussed in \S\ref{sec:velocity_dispersion_caveats}, we cannot definitively conclude that these correlations are significant. However, in case these findings are validated by future larger studies, here we explore the potential implications of this result.

\begin{figure}[h]
\centering
   \includegraphics[angle = 0, clip, width=0.45\textwidth]{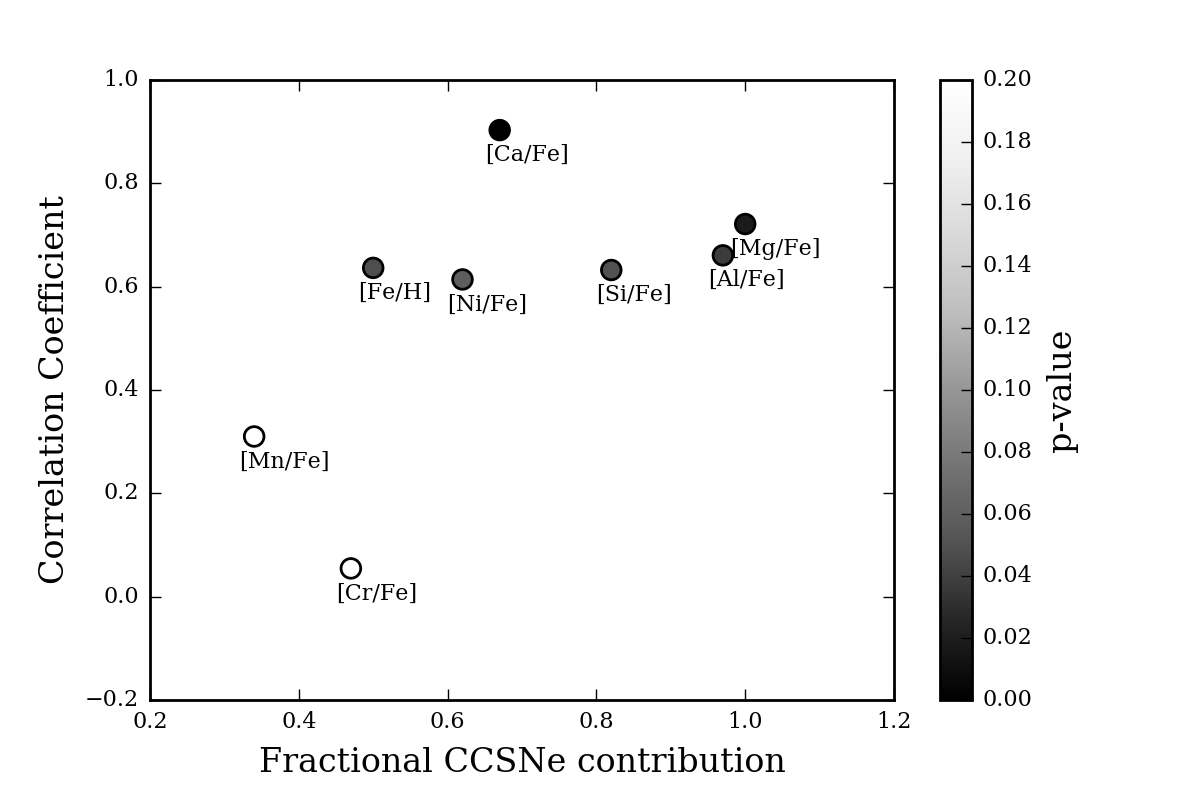}
   \caption{Fractional contribution from CCSNe vs the Spearman correlation coefficient of space velocity dispersion with respect to intrinsic [X/Fe] scatter.}
   \label{fig:abundance_scatter_vs_fcc}
\end{figure}

Figure~\ref{fig:abundance_scatter_vs_fcc} suggests that the nucleosynthetic processes that are responsible for the production of elements in the ISM may have an observable effect on the final abundance scatter within the cluster. It also hints on the existence of a difference in the ejection radii between the pollution mechanisms of Type~Ia SNe and CCSNe. 

Why does the chemical scatter within a cluster depend on the mass of the cluster? Equally interesting, why is this correlation present in certain elements and not in others? The dependence of the intrinsic abundance scatter on cluster mass, can be understood by looking at what we know about cluster formation processes. Clusters are formed from giant molecular clouds in filamentary structures \citep{Kounkel_2019_clusters_filamentary}, forming strings of star-forming gas. So gas accreted to form more massive clusters will not be limited to a sphere surrounding the final cluster, but rather spans a larger range in distance.
\citet{Fujii_2015_YMC_hierarchical_merging} explored the possibility that young massive clusters may be formed by hierarchical merging of sub-clusters or smaller open clusters. 

Although the OCs we are using in our analysis are not as massive as the young massive clusters discussed in \citet{Fujii_2015_YMC_hierarchical_merging}, this mechanism hints that more massive clusters that we study may have been formed by accreting gas over larger ranges in distance in the initial cloud. 
This would result in the more massive clusters to have a larger scatter in metallicity and abundances of certain elements depending on their correlation lengths and mixing efficiency in the initial clouds before star formation began. \citet{Krumholz_2019_Star_clusters_across_cosmic_time} also suggest that massive clusters are formed over extended formation times rather than a single free-fall timescale. This could again increase the chances of the star-forming cloud being polluted by exploding high-mass stars that have already formed.

How are variations in the correlation lengths of different metals in an initial cloud reflected in the chemical homogeneity of the final clusters formed from the cloud? \citet{Armillotta_2018_simulationschemtag} studied this using hydro-dynamical simulations and they observed that if the initial field of the metal is correlated over smaller distances ($<$6~pc), turbulent mixing will efficiently smooth out these inhomogeneities. Metal fields that are correlated over much larger distances ($>$40~pc) will also yield homogeneous stellar chemistry since the variations span a range larger than the typical cloud size. However, metals that are correlated on intermediate lengths in the ISM ($6-40$~pc) can have higher scatter in the stellar abundances of their final clusters. So with this reasoning, elements that have correlation lengths within the intermediate range in the initial cloud may be expected to have higher abundance scatter in the final stellar members for massive clusters. 

Furthermore, since elements belonging to different nucleosynthetic groups have been shown by \citet{Krumholz_2018_metallicityfluctuations} to have different correlation lengths in the initial cloud, we may observe this trend only in certain elements or nucleosynthetic groups. They propose that there should be no significant differences between the correlation lengths of Type~Ia SNe and CCSNe since both types of explosions have comparable energy budgets. 
However, Figure~\ref{fig:abundance_scatter_vs_fcc} suggests the presence of a quantitative difference between the correlation lengths of these two mechanisms that pollute the ISM, and that this difference may manifest itself in the abundance scatter of nucleosynthetic element groups in the final stellar populations.

\section{Summary}
\label{sec:conclusion}

We have identified cluster members for a large number of open clusters in the \citet{Kharchenko_2013_catalog} catalog using only their kinematic information: radial velocities from APOGEE and proper motions from \textit{Gaia}-DR2. 
We provide a catalog of cluster properties and members for 83 clusters with a range of detection quality (\S\ref{sec:validation}).
This cluster membership catalog will be useful for anyone interested in studying cluster chemistry. In addition, we derived new uncertainties for the APOGEE elemental abundances, as a function of stellar parameter and SNR, for the cluster members.

We also studied the dependence of cluster chemical homogeneity on various Galactic and cluster properties. As seen from Figure~\ref{fig:abundance_scatter_vs_dispersion}, Mg and Ca show a strong, relatively significant correlation between cluster chemical scatter and velocity dispersion, while Ni, Si, Al, and Fe may also exhibit a possible positive correlation, albeit at low significance. It is interesting that these elements are those that are predominantly produced by CCSNe. However, we urge caution in these findings due to the small sample size and p-values close to 0.05. Nevertheless, if true, these findings suggest a quantitative difference between the correlation lengths of the metals dispersed into the ISM as a result of Type~Ia SNe and CCSNe, under the assumption that scatter is set by mixing processes. The existence of an intrinsic difference in the distance to which the elements are expelled by these two SNe explosions would affect our understanding of the pollution rates and mixing efficiency in the ISM. For a definite determination, not only is the exploration of more elements required, but also a larger sample of open clusters. If validated by future larger studies, this result should be included in existing and future Galactic chemical evolution models and simulations.

These results also have potential implications for chemical tagging, which first assumes that OCs are intrinsically chemically homogeneous and then attempts to determine birth siblings, cluster members, or co-natal objects using only the chemical signatures of the stars. We find that the abundance scatter in most elements for our clusters are within the limits previously found \citep[e.g.,][]{Bovy_2016_OC_homogeneity}. However, if future studies with a larger OC sample and more elements find similar empirical dependencies of the cluster homogeneity on velocity dispersion, these results should be considered in future work using chemical tagging. For example, the most massive OCs could either be altogether avoided in chemical tagging studies, or be studied with caution for elements that are predominantly produced by CCSNe.

\begin{acknowledgements}
We thank the anonymous referee for thoughtful comments that improved the clarity of the paper.
VP, GZ, KH, and KMK are grateful for support from the Research Corporation for Science Advancement through a Scialog\textsuperscript{\textregistered} award.
SH is supported by an NSF Astronomy and Astrophysics Postdoctoral Fellowship under award AST-1801940. 
JD and PMF acknowledge support for this research from the National Science Foundation (AST-1311835 \& AST-1715662).
DAGH acknowledges support from the State Research Agency (AEI) of the Spanish Ministry of Science, Innovation and Universities (MCIU) and the European Regional Development Fund (FEDER) under grant AYA2017-88254-P.

Funding for the Sloan Digital Sky Survey IV has been provided by the Alfred P. Sloan Foundation, the U.S. Department of Energy Office of Science, and the Participating Institutions. SDSS-IV acknowledges
support and resources from the Center for High-Performance Computing at
the University of Utah. The SDSS web site is www.sdss.org.

SDSS-IV is managed by the Astrophysical Research Consortium for the 
Participating Institutions of the SDSS Collaboration including the 
Brazilian Participation Group, the Carnegie Institution for Science, 
Carnegie Mellon University, the Chilean Participation Group, the French Participation Group, Harvard-Smithsonian Center for Astrophysics, 
Instituto de Astrof\'isica de Canarias, The Johns Hopkins University, Kavli Institute for the Physics and Mathematics of the Universe (IPMU) / 
University of Tokyo, the Korean Participation Group, Lawrence Berkeley National Laboratory, 
Leibniz Institut f\"ur Astrophysik Potsdam (AIP),  
Max-Planck-Institut f\"ur Astronomie (MPIA Heidelberg), 
Max-Planck-Institut f\"ur Astrophysik (MPA Garching), 
Max-Planck-Institut f\"ur Extraterrestrische Physik (MPE), 
National Astronomical Observatories of China, New Mexico State University, 
New York University, University of Notre Dame, 
Observat\'ario Nacional / MCTI, The Ohio State University, 
Pennsylvania State University, Shanghai Astronomical Observatory, 
United Kingdom Participation Group,
Universidad Nacional Aut\'onoma de M\'exico, University of Arizona, 
University of Colorado Boulder, University of Oxford, University of Portsmouth, 
University of Utah, University of Virginia, University of Washington, University of Wisconsin, 
Vanderbilt University, and Yale University.
\end{acknowledgements}

\bibliographystyle{aasjournal}
\bibliography{ms} 

\clearpage

\appendix
\section{}
\label{sec:app_uncertainties}

\begin{figure*}[hbt!]
\includegraphics[angle=0, clip, width=1.0\textwidth]{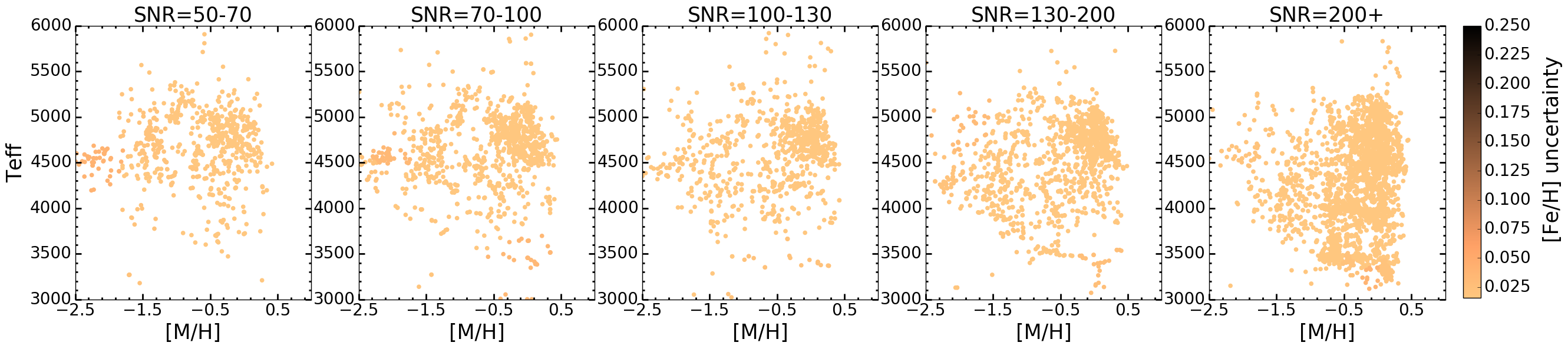}
\includegraphics[angle=0, clip, width=1.0\textwidth]{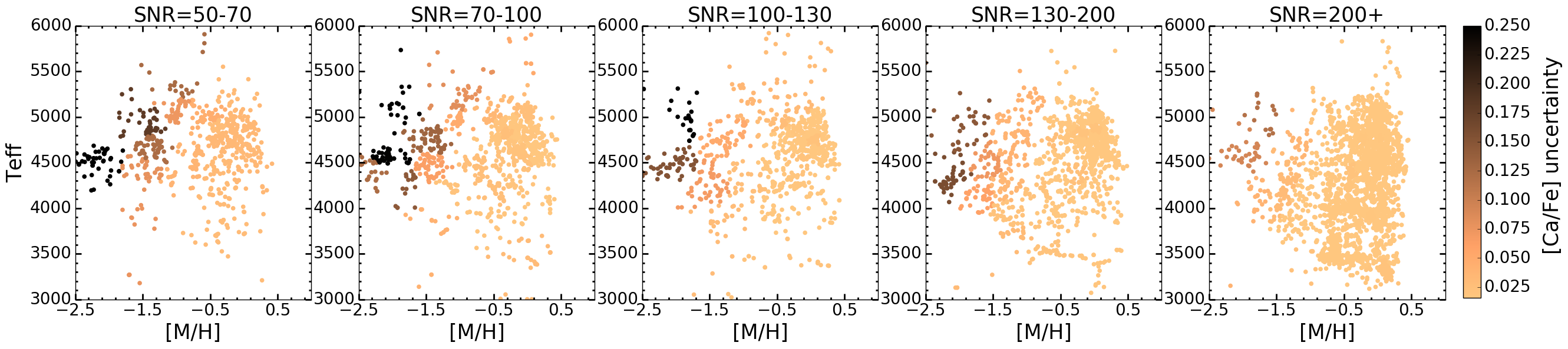}
\includegraphics[angle=0, clip, width=1.0\textwidth]{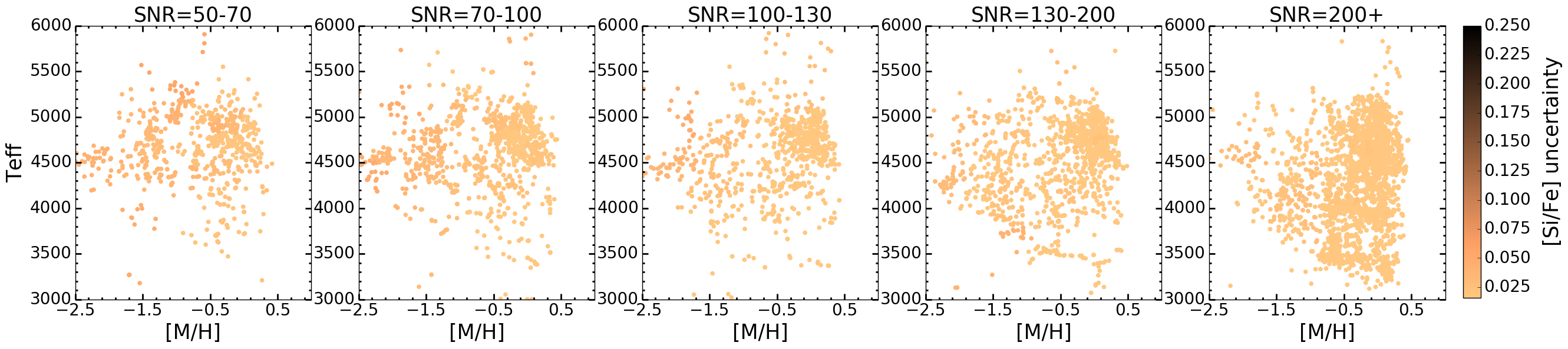} \includegraphics[angle=0, clip, width=1.0\textwidth]{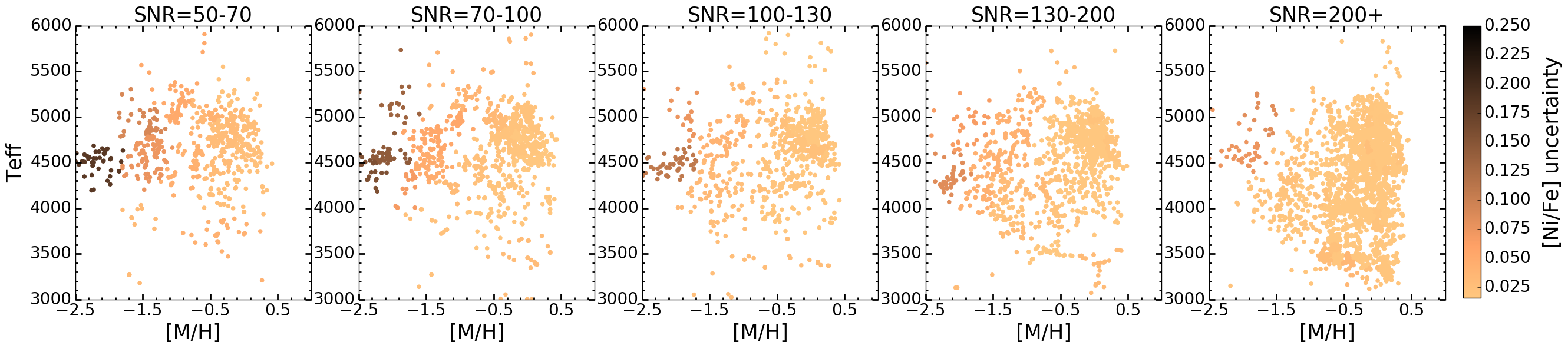} 
\caption{The Uncertainty Training (UT) sample for [Fe/H], [Ca/Fe], [Si/Fe], and [Ni/Fe], similar to first row of Figure \ref{fig:uncertainty_examples}.}
\end{figure*}

\begin{figure*}[hbt!] 
\includegraphics[angle=0, clip, width=1.0\textwidth]{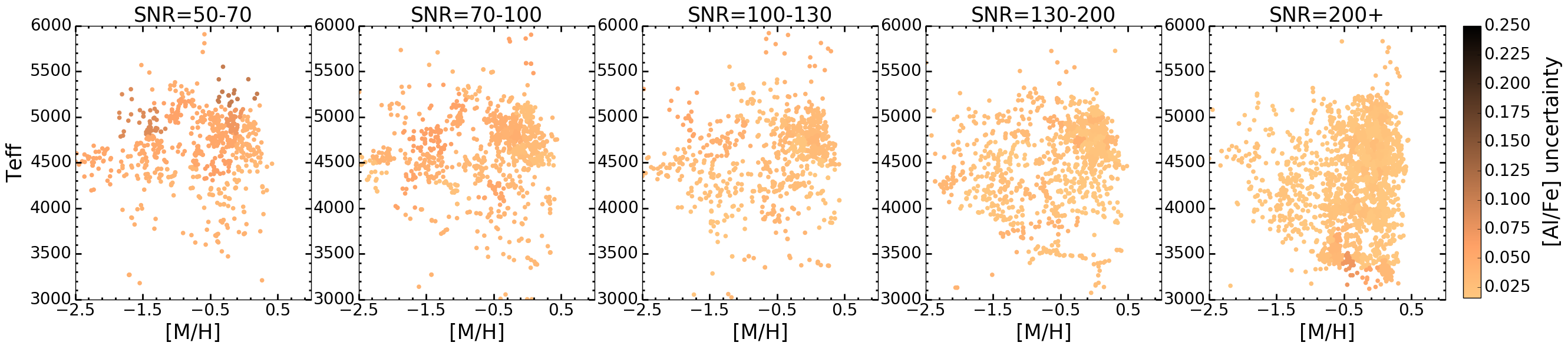} 
\includegraphics[angle=0, clip, width=1.0\textwidth]{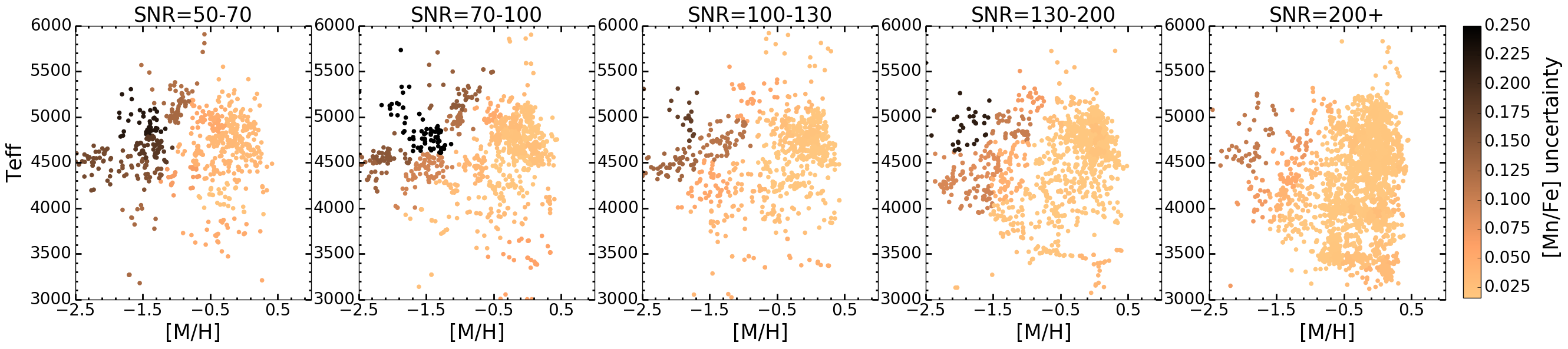} 
\includegraphics[angle=0, clip, width=1.0\textwidth]{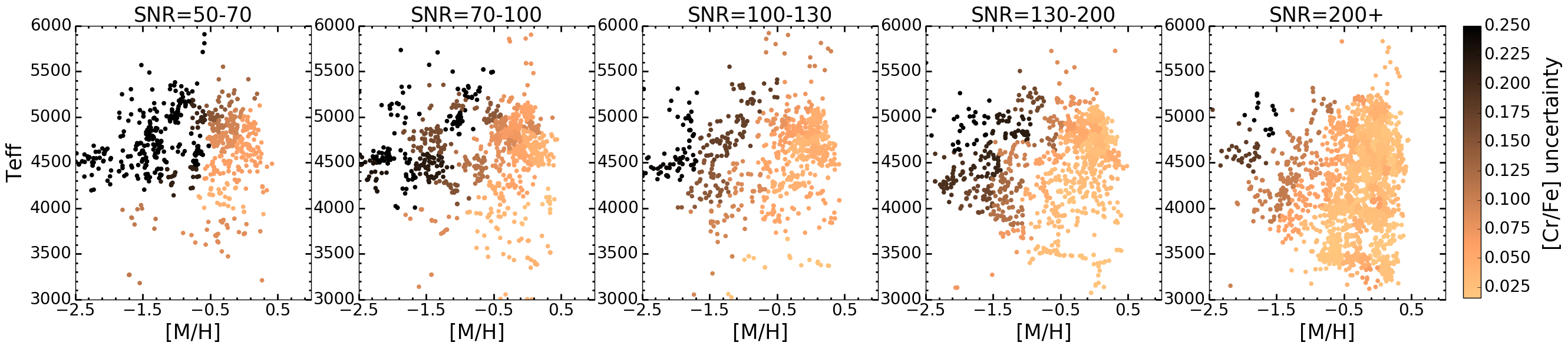} 
\caption{The Uncertainty Training (UT) sample for [Al/Fe], [Mn/Fe], and [Cr/Fe], similar to first row of Figure \ref{fig:uncertainty_examples}.}
\end{figure*}

\clearpage

\newpage

\section{}
\label{sec:app_membership}

Here we show the membership plots for all of the clusters that we use in the final analysis in \S\ref{sec:results}:

\begin{figure}[ht]
\begin{minipage}[b]{0.5\textwidth}
\includegraphics[width=0.8\textwidth]{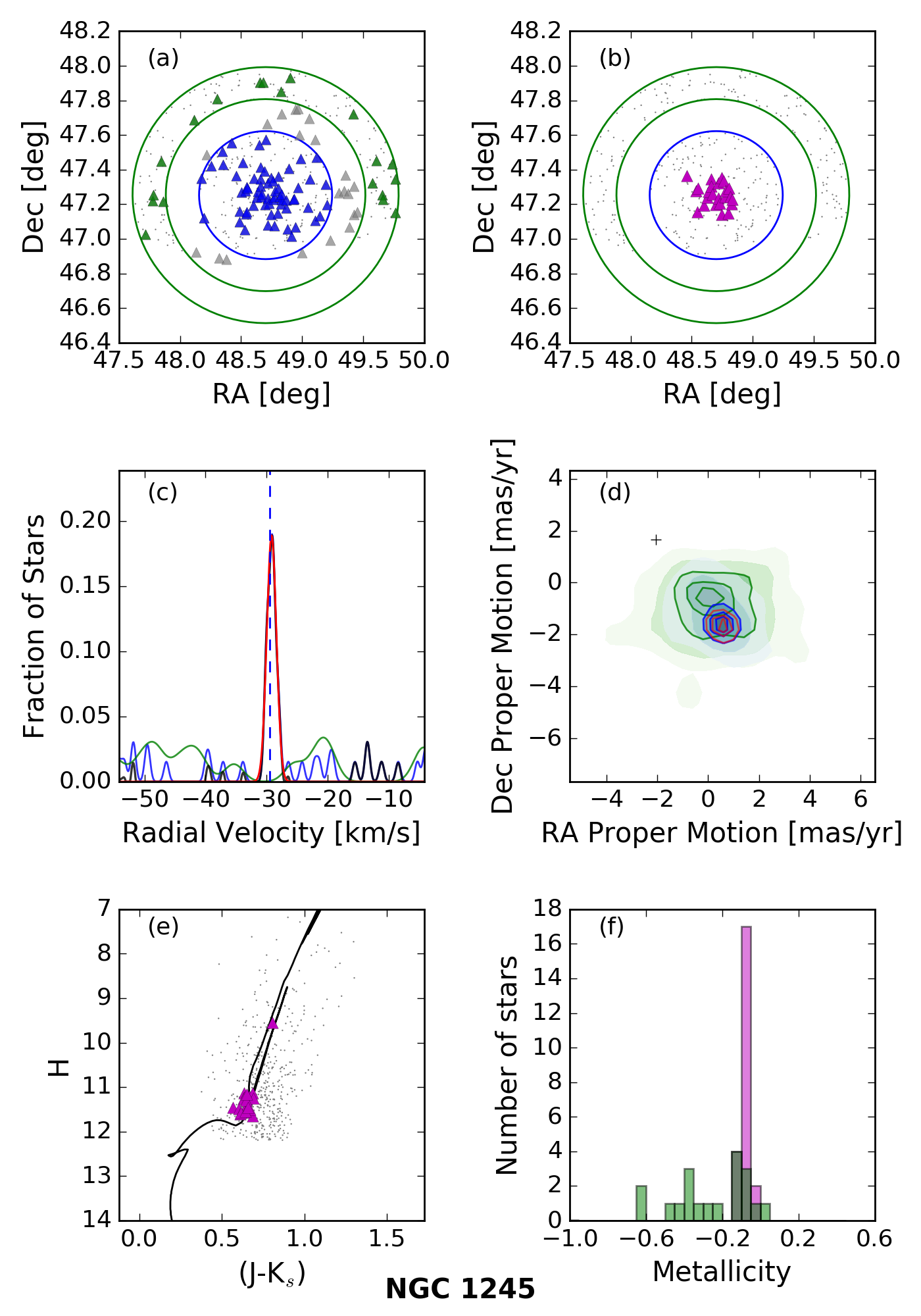} 
\caption{
Same as Figure \ref{fig:membership_NGC6819}, but for NGC 1245.
}
\end{minipage}
\begin{minipage}[b]{0.5\textwidth}
\includegraphics[width=0.8\textwidth]{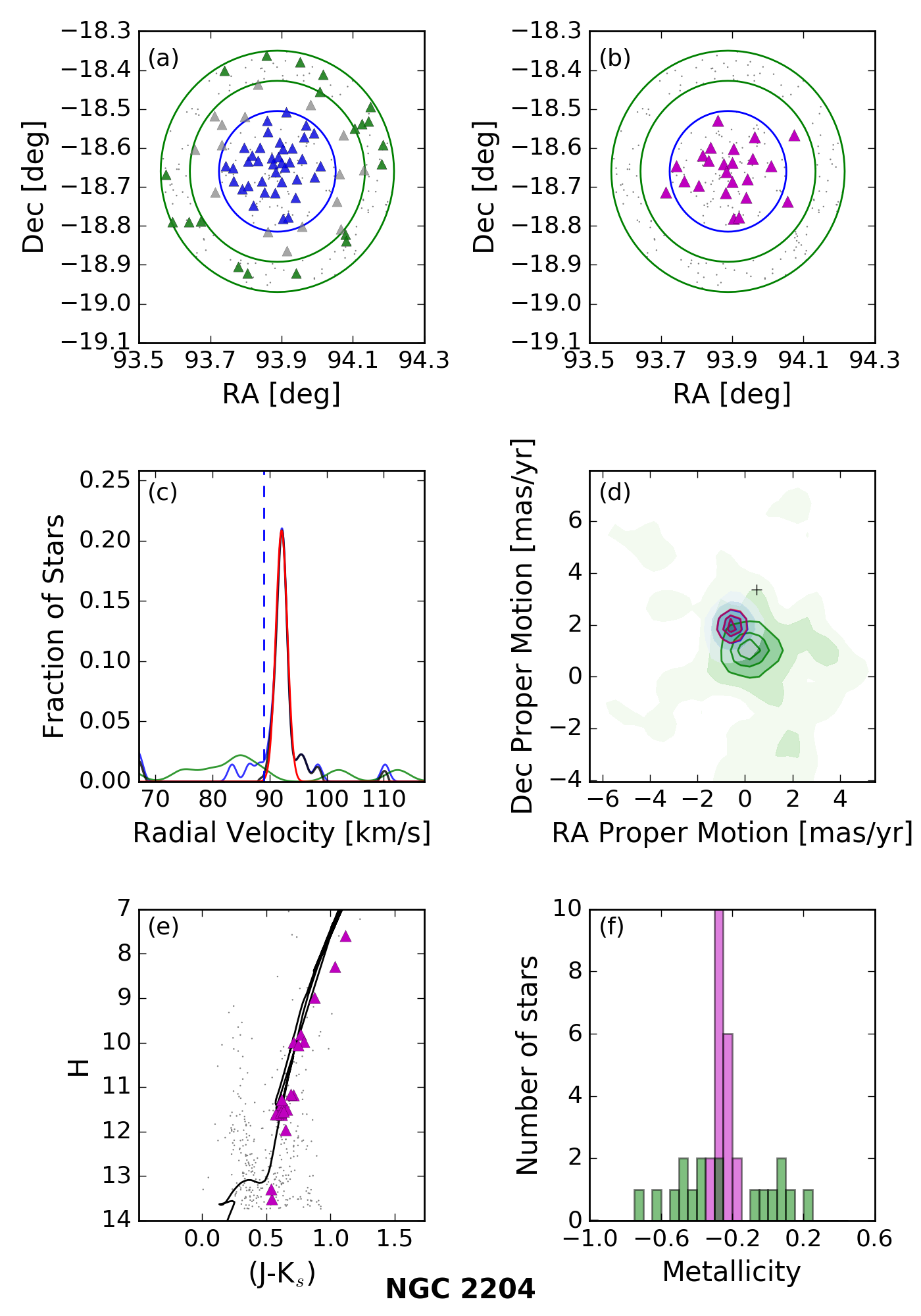} 
\caption{
Same as Figure \ref{fig:membership_NGC6819}, but for NGC 2204.
}
\end{minipage} 
\begin{minipage}[b]{0.5\textwidth}
\includegraphics[width=0.8\textwidth]{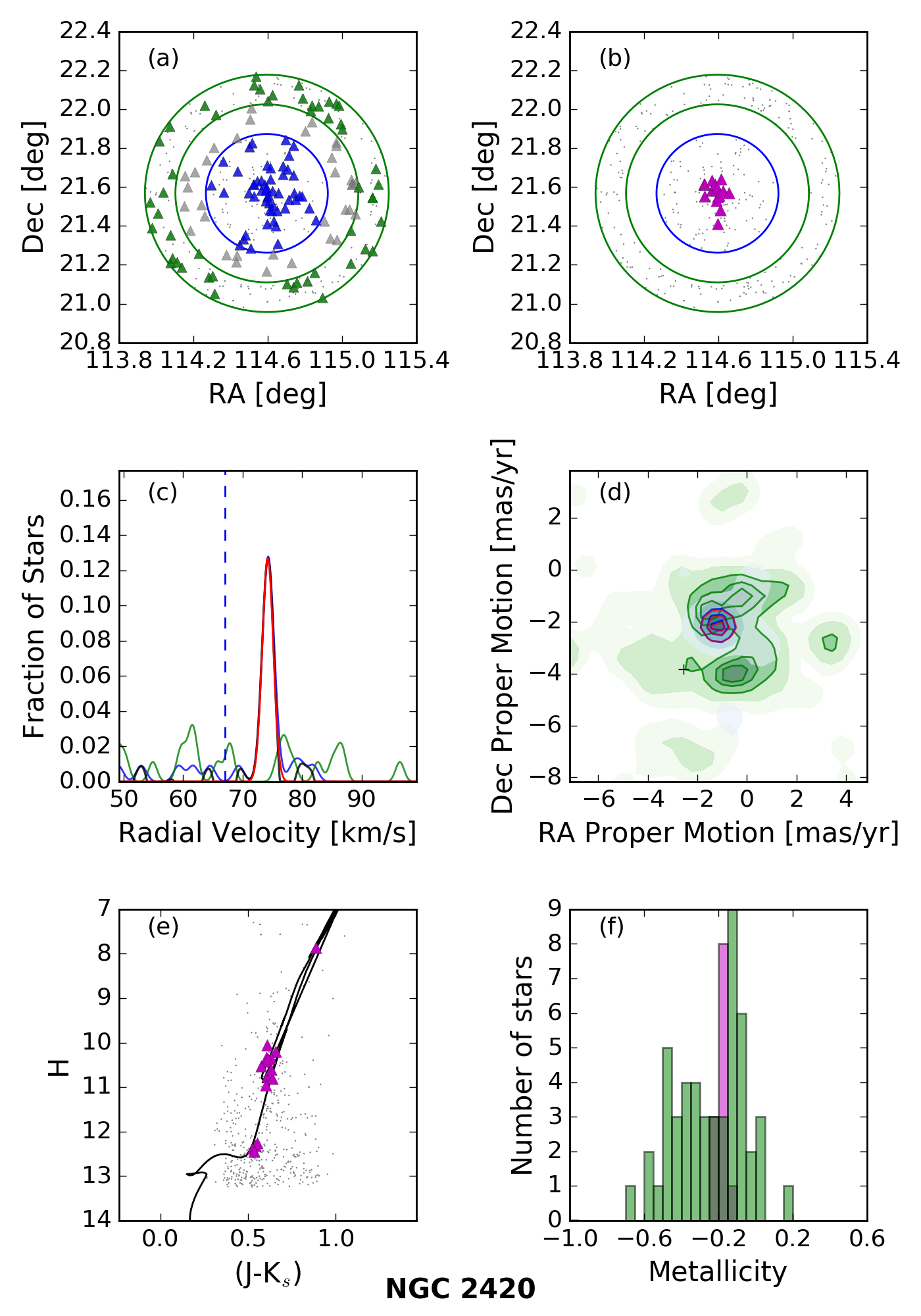} 
\caption{
Same as Figure \ref{fig:membership_NGC6819}, but for NGC 2420.
}
\end{minipage} 
\begin{minipage}[b]{0.5\textwidth}
\includegraphics[width=0.8\textwidth]{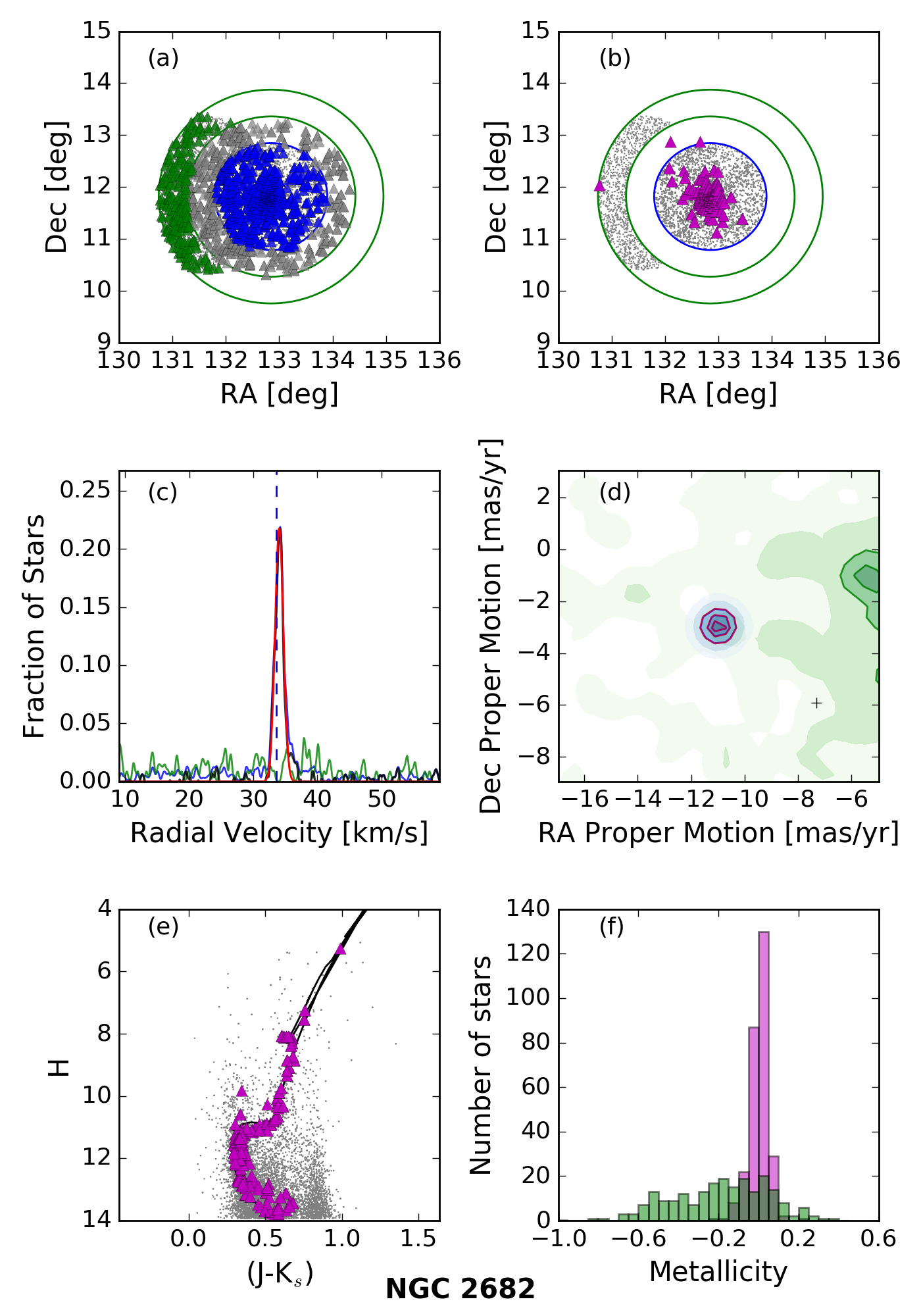} 
\caption{
Same as Figure \ref{fig:membership_NGC6819}, but for NGC 2682.
}
\end{minipage}
\end{figure} 

\begin{figure}[ht]
\begin{minipage}[b]{0.5\textwidth}
\includegraphics[angle=0, clip, width=0.8\textwidth]{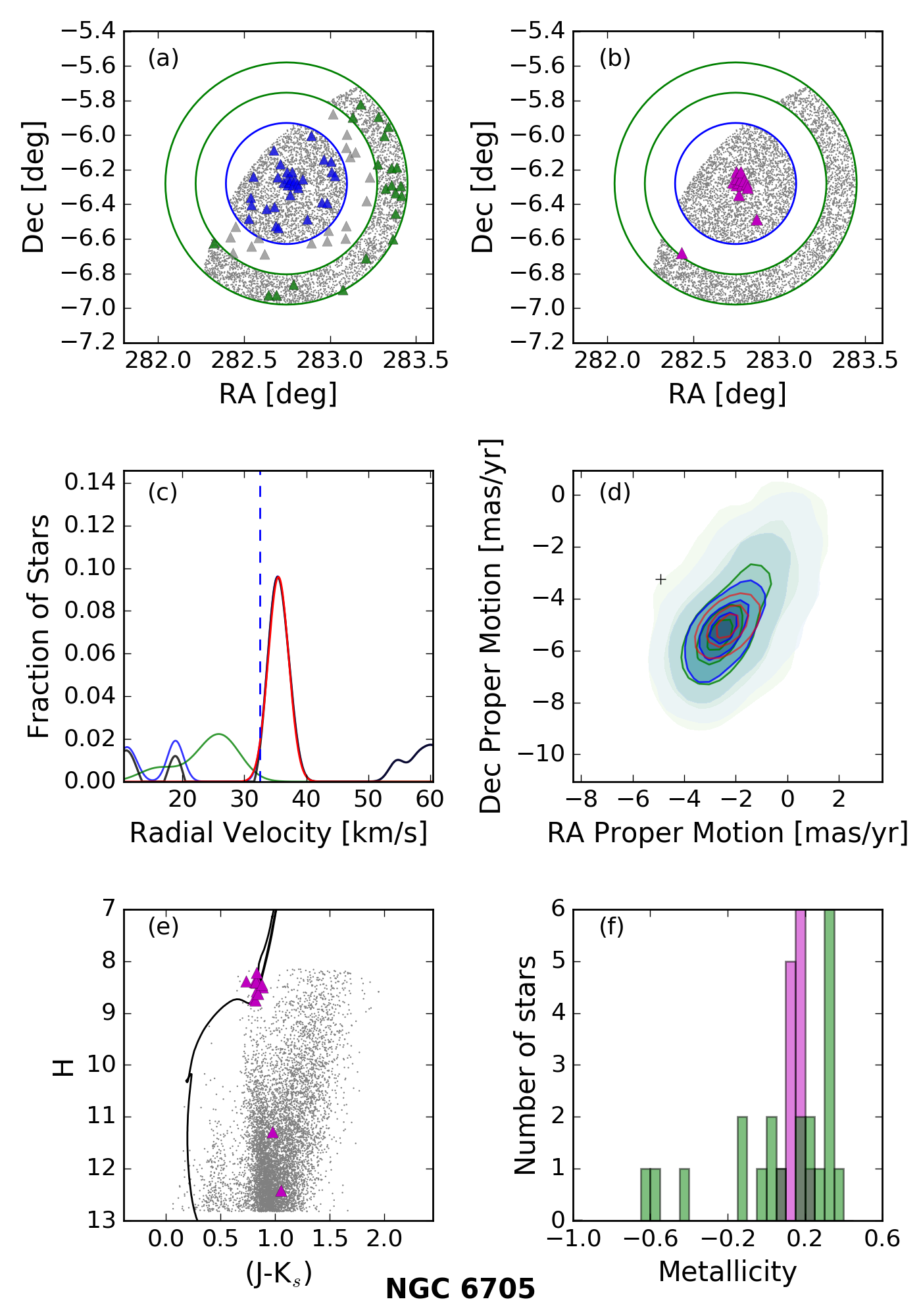} 
\caption{
Same as Figure \ref{fig:membership_NGC6819}, but for NGC 6705.
}
\end{minipage} 
\begin{minipage}[b]{0.5\textwidth}
\includegraphics[angle=0, clip, width=0.8\textwidth]{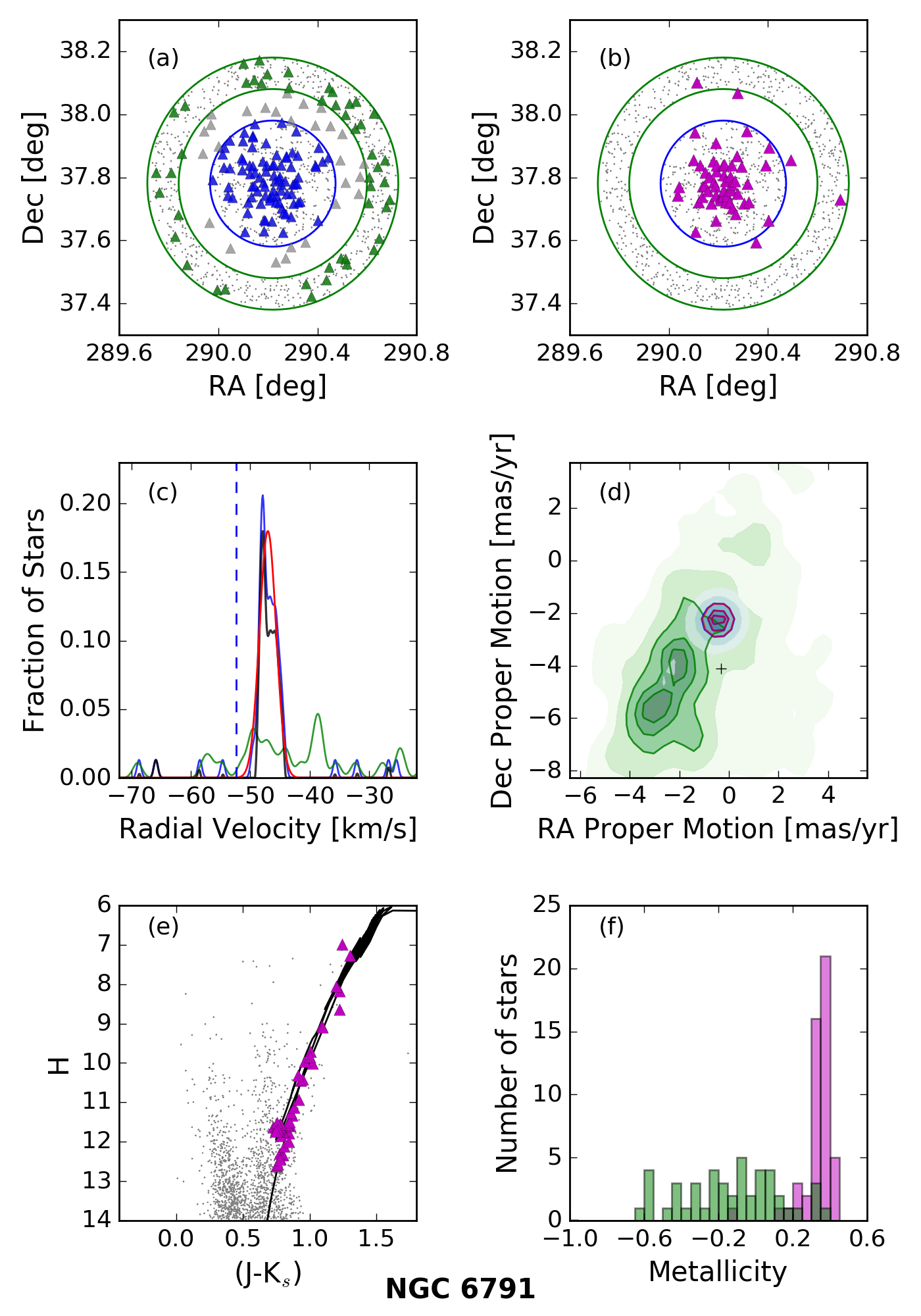} 
\caption{
Same as Figure \ref{fig:membership_NGC6819}, but for NGC 6791.
}
\end{minipage} 
\begin{minipage}[b]{0.5\textwidth}
\includegraphics[angle=0, clip, width=0.8\textwidth]{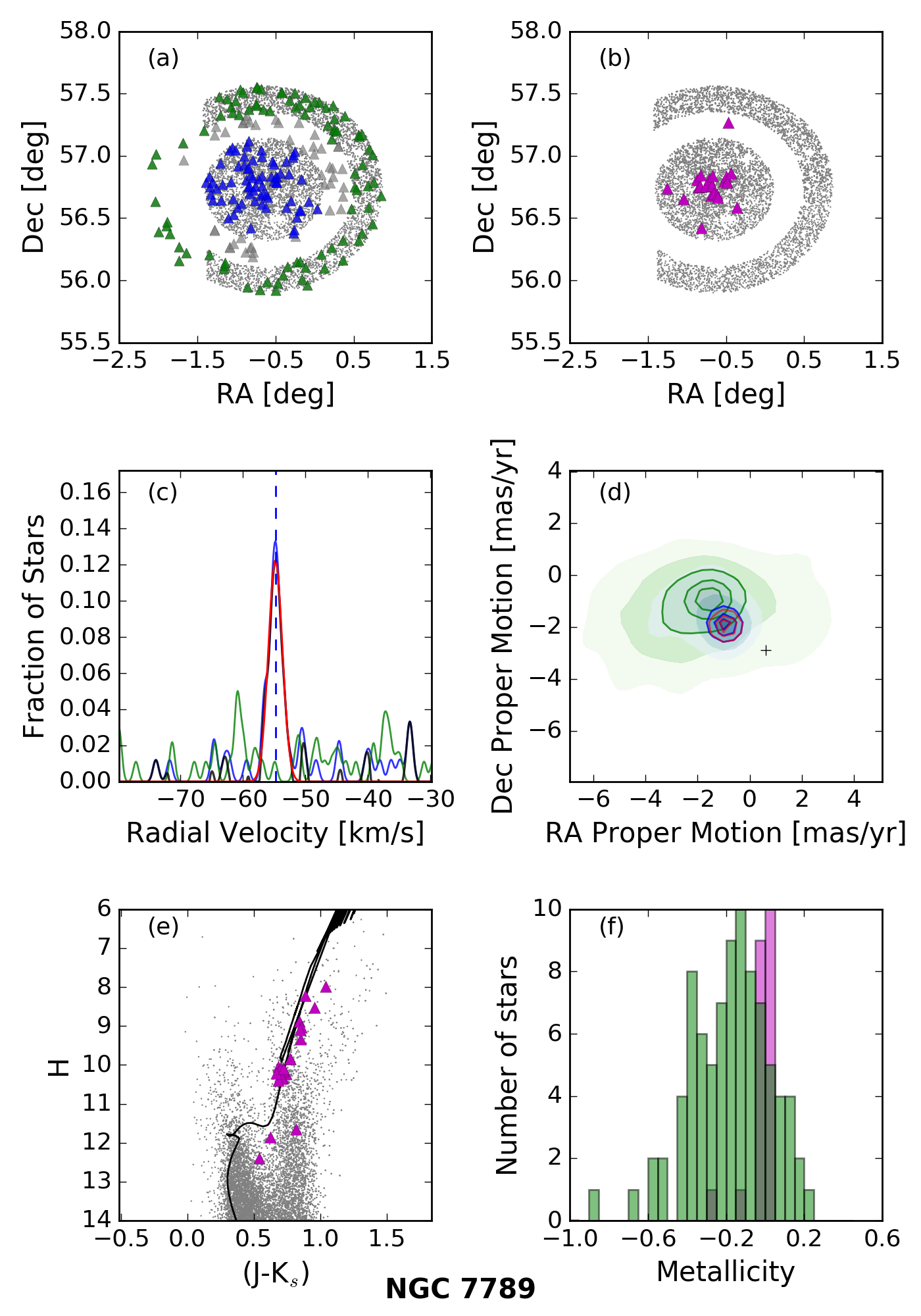} 
\caption{
Same as Figure \ref{fig:membership_NGC6819}, but for NGC 7789.
}
\end{minipage} 
\begin{minipage}[b]{0.5\textwidth}
\includegraphics[angle=0, clip, width=0.8\textwidth]{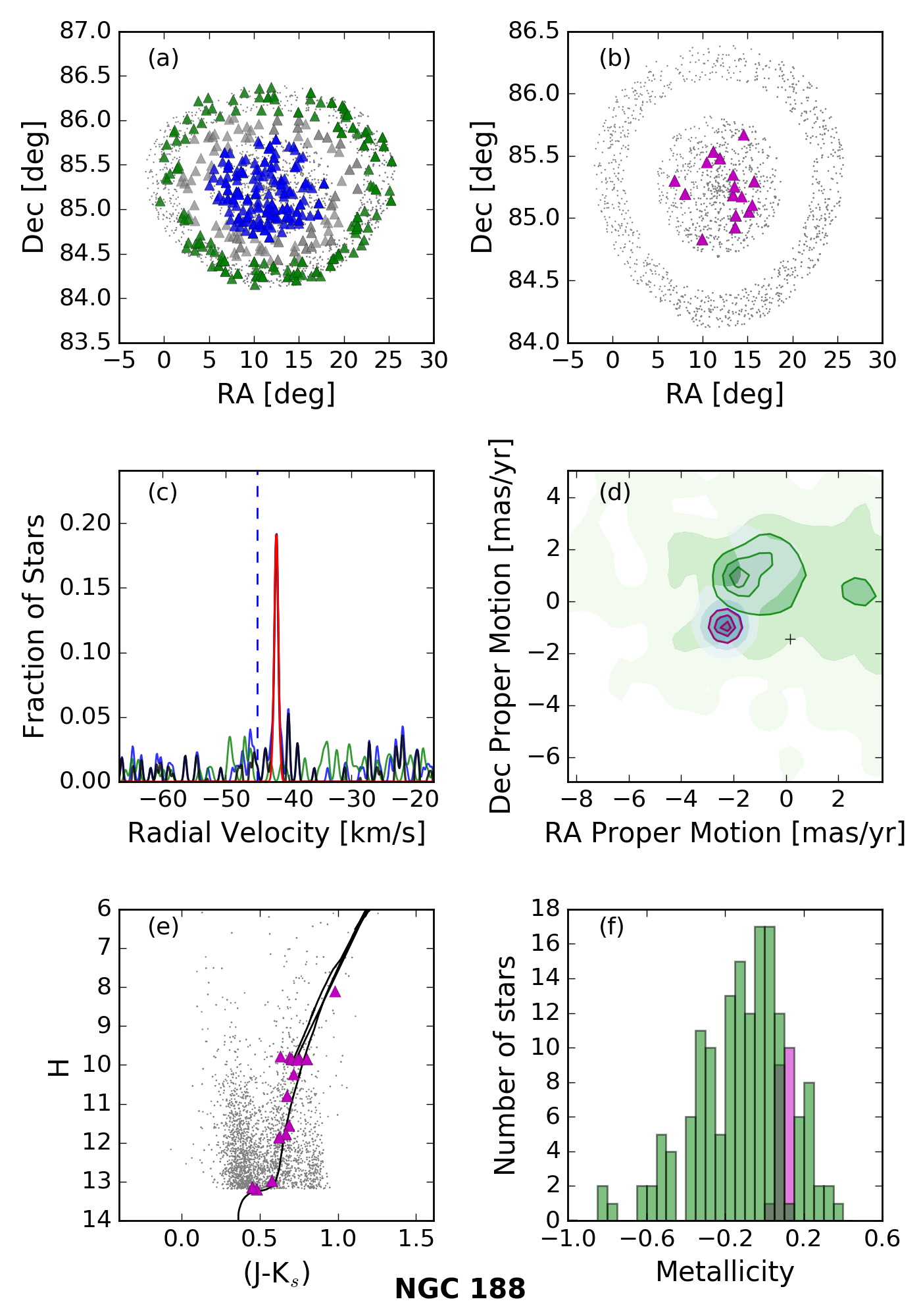} 
\caption{
Same as Figure \ref{fig:membership_NGC6819}, but for NGC 188.
}
\end{minipage} 
\end{figure} 

\begin{figure}[ht]
\begin{minipage}[b]{0.5\textwidth}
\includegraphics[angle=0, clip, width=0.8\textwidth]{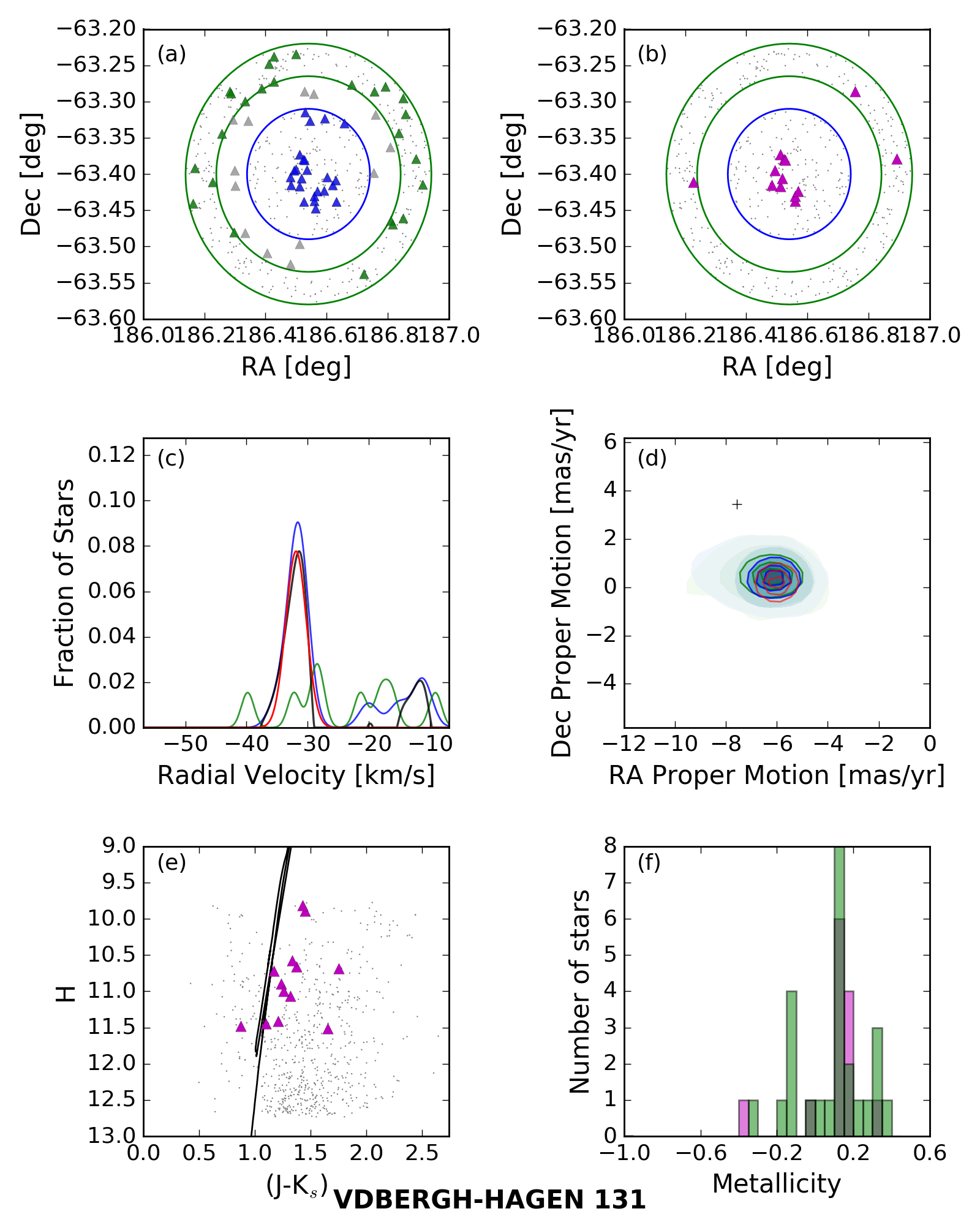} 
\caption{
Same as Figure \ref{fig:membership_NGC6819}, but for VDBERGH-HAGEN 131.
}
\end{minipage} 
\end{figure}

\end{document}